\documentclass[usenatbib]{mn2e} 
\usepackage{url}
\usepackage{amssymb}
\usepackage{graphicx}
\usepackage{abbreviations} 
\footnotesize 
\newdimen\digitwidth %define! a one digit width for tables 
\setbox0=\hbox{\rm0} 
\digitwidth=\wd0
\catcode`!=\active 
\def!{\kern\digitwidth} 
\normalsize

\title[DM Variations and Pulsar Timing]{Dispersion Measure Variations
and their Effect on Precision Pulsar Timing}

\author[X. P. You et al.]  {X. P. You,$^{1,2,3}$\thanks{Email:
xpyou@bao.ac.cn} G. Hobbs,$^2$ W. A. Coles,$^{4,2}$ R. N. Manchester,$^2$
R. Edwards,$^2$ M. Bailes,$^5$ 
\newauthor J. Sarkissian,$^2$  J. P. W. Verbiest,$^{5,2}$ W. van Straten,$^6$ 
A. Hotan,$^7$ S. Ord,$^8$ 
\newauthor F. Jenet,$^6$ N. D. R. Bhat,$^5$ A. Teoh$^{5,2}$ \\
$^1$ National Astronomical Observatories, Chinese Academy of
Sciences, Beijing 100012, China \\ 
$^2$ Australia Telescope National Facility, CSIRO, PO~Box~76, Epping, 
NSW~1710, Australia \\ 
$^3$ Graduate University of Chinese Academy of Sciences, Beijing 100049, China \\
$^4$ Electrical and Computer Engineering, University of California at San
Diego, La Jolla, California, USA\\ 
$^5$ Centre for Astrophysics and Supercomputing, Swinburne University of 
Technology, P. O. Box 218, Hawthorn VIC 3122, Australia \\ 
$^6$ Center for Gravitational Wave
Astronomy, University of Texas at Brownsville, TX 78520 \\
$^7$ School of Mathematics and Physics, University of Tasmania, 
Hobart TAS 7001, Australia \\
$^8$ School of Physics, University of Sydney, NSW 2006, Australia
}

\let\leqslant=\leq %Authors with AMS fonts and mssymb.tex can comment
\date{}
\begin{document}
\maketitle
\newcommand{\setthebls}{
%                 de-comment this line for double spacing:
%\baselineskip=20pt
}
\setthebls
\begin{abstract}
We present an analysis of the variations seen in the dispersion
measures (DMs) of 20 millisecond pulsars observed as part of the
Parkes Pulsar Timing Array project. We carry out a statistically
rigorous structure function analysis for each pulsar and show that the
variations seen for most pulsars are consistent with those expected
for an interstellar medium characterised by a Kolmogorov turbulence
spectrum. The structure functions for PSRs~J1045$-$4509 and
J1909$-$3744 provide the first clear evidence for a large inner scale,
possibly due to ion-neutral damping.  We also show the effect of the
solar wind on the DMs and show that the simple models presently
implemented into pulsar timing packages cannot reliably correct for
this effect.  For the first time we clearly show how DM variations
affect pulsar timing residuals and how they can be corrected in order
to obtain the highest possible timing precision. Even with our
presently limited data span, the residuals (and all parameters derived
from the timing) for six of our pulsars have been significantly
improved by correcting for the DM variations.
\end{abstract}

\section{Introduction}

The Parkes Pulsar Timing Array (PPTA) is a project which aims to take
advantage of the extraordinary rotational stability of short period
(millisecond) radio pulsars. The principal goal of the PPTA is to make
a direct detection of gravitational waves \citep{hob05,man06}.  For
this purpose it is necessary to measure weekly times of arrival (TOAs)
of $\sim$20 pulsars with a precision between 100 and 500\,ns
\citep{jhlm05}.  In order to achieve this goal all systematic errors
in the TOAs must be considered and, if possible, corrected.  One such
correction is the delay caused by the plasma between the pulsar and
the Earth.  Most of this plasma contribution is from the interstellar
medium, but the contribution of the solar wind cannot be neglected and
the ionosphere will occasionally be important.  The dispersion in the
plasma is a linear effect and can, in principle be corrected exactly.
The group delay, $t_g(\nu)$, is related to the integral of the
electron density, $n_e$, from the Earth to the pulsar, $t_g(\nu) =
{\rm DM}/(K\nu^2)$, where
\begin{equation}
 \mbox{DM} = \int_0^L n_{e} \mbox{\ d} \mathnormal{l},
\end{equation}
is the ``dispersion measure''.  The dispersion constant $K \equiv
2.410\times10^{-4}$\,MHz$^{-2}$\,cm$^{-3}$\,pc\,s$^{-1}$, $\nu$ is the
observing frequency and $L$ the distance from the Earth to the pulsar.
When TOAs, $t_{g1}$ and $t_{g2}$, are measured at two frequencies, $\nu_1$
and $\nu_2$, the DM can be estimated using
\begin{equation}
\mbox{DM} = K \frac{t_{g2} - t_{g1}}{\nu_2^{-2} - \nu_1^{-2}}.
\label{eq:dm1}
\end{equation}

A rough estimate of the DM of a (radio) pulsar is generally obtained
from the discovery observations. This estimate can be quickly refined
by re-observing the pulsar with more widely separated
frequencies. Measured DMs for currently known radio pulsars lie
between 2.38 and 1235\,cm$^{-3}$\,pc and from 2.65 to 244\,cm$^{-3}$\,pc
for the subset of millisecond pulsars
\citep{mhth05}\footnote{http://www.atnf.csiro.au/research/pulsar/psrcat}.

Precise measurements of DM show that it often has significant time
variations.  A time delay of 100\,ns at an observing frequency of
1400\,MHz, the accuracy goal of the PPTA, corresponds to a DM
variation of $4.72\times10^{-5}$\,cm$^{-3}$\,pc.  Variations of this
order can occur in the ionosphere only for zenith angles in excess of
80$^\circ$ or during major geomagnetic storms, so ionospheric
corrections will not normally be necessary. At this level of timing
precision, significant variations in DM can occur due to the solar
wind even when the pulsar is 60$^{\circ}$ away from the
Sun. Variations in the interstellar plasma DM result from plasma
turbulence and usually have a Kolmogorov power spectrum, implying that
the variations are larger over longer time-scales.  In the pulsars
observed by the PPTA project, such DM fluctuations can reach levels
that require correction within a few days or weeks.

It is clear that the goals of the PPTA project cannot be reached
without measuring the DM and correcting for the plasma delay for each
observation.  The most precise TOA measurements are usually obtained
at a frequency of 1400\,MHz, but the only dual-band receiver available
at the Parkes telescope is at 685 and 3100\,MHz.  Thus the DM
variations are measured using the dual-band system at different times
than the TOA observations at 1400\,MHz.  Since the DM varies
relatively smoothly, the DM correction can be interpolated to the
epoch of the primary TOA observation.  In this paper we use the first
few years of DM measurements to test methods of correcting for the
solar wind, to study the interstellar plasma turbulence and to derive
algorithms for correcting the TOA measurements.

\section{Causes of DM variations}

The contributions of the ionosphere and the solar wind have been
well-studied and can be estimated by various methods independently of
the PPTA.  The total electron content of the ionosphere (``TEC'') is
regularly monitored because it is needed to correct the Global
Positioning System (GPS) navigational system. A monitor is located at
the Parkes observatory, but it is seldom necessary to make this
correction.  Corrections for the solar wind are implemented in the
standard pulsar timing codes \textsc{tempo} and \textsc{tempo2}
\citep{hem06,ehm06}. However, these assume a spherically symmetric
solar wind with a constant scale factor and do not model observed
variations in wind density with latitude, longitude and time which can
be as much as a factor of four at any distance \citep{mbf+00}. The
earlier package, \textsc{tempo}, assumes a higher density compared to
\textsc{tempo2}. Neither of these is adequate for the desired PPTA
precision.

The DM due to the interstellar medium varies for a variety of reasons.
For example, variations are known to occur for some pulsars within
supernova remnants, when wisps of ionised gas drift across the line of
sight to the pulsar.  For instance, the DM of the Vela pulsar
decreased at a rate of 0.04\,cm$^{-3}$\,pc\,yr$^{-1}$ from 1970 to
1985 \citep{hhc85}.  Similarly the Crab pulsar shows variations up to
0.02\,cm$^{-3}$\,pc\,yr$^{-1}$ over 15\,yr \citep{lps88}.  Pulsars in
binary systems which exhibit eclipses show DM variations from the
ionised envelope of the companion object. These have been measured for
two of the binary pulsars in the globular cluster 47 Tucanae
\citep{fck+03}.  The DM change of 0.0065\,cm$^{-3}$\,pc for one of
these, PSR~J0023$-$7203J, is 100 times the level that would require
correction for the PPTA pulsars.  Even larger changes have been
observed in PSR~B1259$-$63 which is in orbit with a massive B2e star,
reaching 10.7 and 7.7\,cm$^{-3}$\,pc during the periastron passages 
of 1994 and 1997, respectively \citep{wjm04}.

The DM also varies due to turbulent spatial variations which drift
across the line of sight between the Earth and the pulsar.  These have
commonly been characterised in the literature as linear slopes in DM.
Measurements of such $\mbox {dDM}/ \mbox{d}t$ values for four pulsars
were discussed by \citet{bhvf93} who proposed that $\mbox {dDM}/
\mbox{d}t \propto ({\rm DM})^{1/2}$ and modelled the variation using
wedges of enhanced density.  Observations of 374 pulsars were
presented by \citet{hlk+04} who found the same relationship.  However,
a better characterisation of the DM variations can be made using the
theoretical spatial characteristics of a turbulent process. As shown
later, in a turbulent model the relation $\mbox {dDM}/ \mbox{d}t
\propto ({\rm DM})^{1/2}$ arises naturally and does not require a
wedge model. The spatial power spectrum of electron density was
defined by \citet{ric90} as
\begin{eqnarray}
P(q)=C_n^2 q^{-\beta};\;\;  \quad 2\pi/l_o < q < 2\pi/l_i,
\end{eqnarray}
where $C_n^2$ scales the power spectrum (and thus the total energy in
the process), $\beta$ is the power-law exponent (which is $11/3$ for a
Kolmogorov spectrum), $l_o$ is the outer scale and $l_i$ is the inner
scale.  Physically the outer scale is identified with the largest
scale in the medium, typically the size at which it becomes
inhomogeneous, and the inner scale is the scale at which dissipation
occurs. Energy is introduced at some scale between $l_o$ and $l_i$,
supporting the spectrum.  This energy `cascades' in frequency to $l_i$
where it is dissipated. Turbulent variations in electron density can
be estimated from DM variations and various diffractive phenomena such
as angular scattering, pulse broadening and intensity scintillations.
Diffractive variations are caused by much smaller scale fluctuations
in density and thus probe different regions of the spatial spectrum
than DM variations.  Diffractive variations are modulated by
refractive variations which can be used to probe scales between the
diffractive and the DM scales.  All these observed variations result
from the motion of the line of sight through the scattering medium,
thus spatial variations of scale $s$ are associated with temporal
variations of scale $T$ by $s = VT$ where $V$ is the velocity of the
line of sight with respect to the scattering plasma. Therefore, the
inner time-scale $\tau_i$ corresponds to $l_i = V\tau_i$.

Radio scattering observations, such as those mentioned above, are
directly sensitive to a statistic called the ``phase structure
function'', $D_\phi(\tau)$, which is defined by
\begin{equation}
D_\phi(\tau) = \langle[\phi(t+\tau) - \phi(t)]^2\rangle,
\label{eq:stdiff}
\end{equation}
where $\phi$ is the geometrical phase delay between the source and the
observer and the angle brackets denote an ensemble average.  For a
power-law spatial spectrum with $2<\beta<4$ between the inner and
outer scales, the structure function $D_{\phi}$ is given by
\citep{ric77}:
\begin{eqnarray}
D_{\phi}(\tau)=\left(\tau/\tau_d\right)^{\alpha},
\label{eq:sts}
\end{eqnarray}
where $\alpha = \beta - 2$.  Similarly, we can define a structure
function for the DM variations:
\begin{equation}
D_{\rm DM}(\tau) = \langle[\mbox{DM}(t+\tau) - \mbox{DM}(t)]^2\rangle.
\label{eq:stdm}
\end{equation}
At scales that are larger than the scale of refractive scintillation,
the geometrical phase approaches the actual phase \citep{clhm91} and
these two structure functions can be related through the dispersion
relation (Equation~\ref{eq:dm1}),
\begin{equation}
D_{\rm DM}(\tau) = \left(K \nu/2\pi\right)^2 D_\phi(\tau).
\label{eq:stdphi}
\end{equation}

The structure function was first used to investigate DM variations by
\citet{ric88}. Subsequently, the technique was applied to
PSR~J1939$+$2134 (B1937$+$21) by \citet{cwd+90}, \citet{ktr94} and
\citet{rdb+06}.  \citet{cwd+90} showed that the structure function was
consistent with a power-law fluctuation spectrum with index $\beta$
between 11/3, the Kolmogorov value, and 4. \citet{ktr94} continued
this work and obtained $\beta = 3.874\pm 0.011$. From the approximate
agreement of the diffractive timescale $\tau_d$ computed from
Equation~\ref{eq:sts} and the directly measured value, \citet{cwd+90}
inferred that the inner scale of the fluctuation spectrum, $l_i$, was
less than about $2\times10^{7}$\,m.  Recently, \citet{rdb+06} extended
the data-span to 20\,yr and obtained $\beta = 3.66 \pm 0.04$ which is
consistent with the value expected for a Kolmogorov spectrum and
suggested $l_i \sim 1.3\times 10^9$\,m. \citet{cl97a} presented the DM
variations of a different millisecond pulsar, PSR~J1824$-$2452
(B1821$-$24), and obtained $\beta = 3.7 \pm 0.2$ which is also
consistent with a Kolmogorov spectrum. Dispersion measure variations
were measured for six pulsars by \citet{pw91} and structure functions
were obtained for PSRs B0834+06, B0823+26 and B0919+06. Measured
power-law indices were in the range 3.77 to 3.87 with uncertainties of
0.04 or less.

Assuming that the DM variations are due to turbulence then, from the
definition of the structure function, the ``slope'' $\mbox {dDM}/
\mbox{d}t$, averaged over an interval $\tau$, will be a random
variable with an rms of $[D_{\rm DM}(\tau)]^{1/2}/\tau$. This can be
related to the mean DM value by the distance to the pulsar $L$ as both
DM and $D_{\rm DM}(\tau)$ are proportional to $L$. Thus the observed
result that $\mbox {dDM}/\mbox{d}t \propto ({\rm DM})^{1/2}$ is
expected for any turbulent medium and does not require ad hoc models
such as the wedge model of \citet{bhvf93}. This proportionality will
be valid for spatial scales $V\tau$ that are less than the parsec
scale of interstellar clouds, since it assumes that contributions to
the DM fluctuations from various points on the line of sight add
incoherently.

There have been three dissipation mechanisms discussed in the
literature: ion cyclotron damping (which is the primary mechanism in
the solar wind), Landau damping and ion-neutral collisional damping.
It is not thought that Landau damping is important in the ISM
\citep{ms97}.  Ion cyclotron damping will certainly occur if the
turbulent cascade reaches the small spatial scales involved. It occurs
at the ion inertial scale \citep{ch89},
\begin{equation}
L_i = 684$\,km$/\sqrt{n_e({\rm cm}^{-3})} 
\label{eqn:li}
\end{equation}
and has been clearly observed in the solar wind. Expected scales in
the ISM range from 300 to 3000\,km and it has almost certainly
been observed at scales of 300 to 800\,km using pulse broadening
observations \citep{bcc+04}.  Damping due to ion-neutral collisions is
also a resonant process that occurs near the ion-neutral collision
frequency and would result in scales of $\sim$30\,AU in typical warm
ISM conditions \citep{ms97}.

\section{Observations and data analysis}

The PPTA, which commenced observations in February 2004, uses the
Parkes 64-m radio telescope to make timing observations of 20
millisecond pulsars.  One, PSR~J1824$-$2452, lies within the globular
cluster M28, the others within the disk of our
Galaxy. Table~\ref{tb:sum} lists basic parameters for the 20 PPTA
pulsars: J2000 name, period ($P$), period derivative ($\dot{P}$),
orbital period ($P_b$) if the pulsar is in a binary system, DM,
distance based on the NE2001 electron density model \citep{cl02b}
unless the annual parallax or another independent distance estimate is
available, total proper motion ($\mu$) and ecliptic latitude
($b_E$). Because PSRs~J1022$+$1001 and J1730$-$2304 lie very close to
the ecliptic plane, timing methods cannot provide a precise proper
motion in ecliptic latitude; for these two pulsars the proper motion
in ecliptic longitude is given. Each pulsar is typically observed at
intervals of 2--3 weeks at frequencies close to 685\,MHz (50\,cm),
1400\,MHz (20\,cm) and 3100\,MHz (10\,cm), where the band designations
(based on wavelength) are given in parentheses. We have used three
backend systems: a wideband correlator (WBC), a digital filterbank
system (DFB1)\footnote{A new digital filterbank system (DFB2) with a
wider bandwidth and improved resolution is currently under
construction.} and the Caltech-Parkes-Swinburne Recorder 2 (CPSR2), a
coherent dedispersing system, all of which record orthogonal linear
polarisations. The WBC provides 2-bit sampling with a bandwidth of up
to 1024\,MHz for the earlier data. The DFB1 was installed in 2005 June
and allows 8-bit sampling of a 256 MHz bandwidth at 10\,cm and
20\,cm. Observations at 10 and 50\,cm are obtained simultaneously
using a dual-band receiver providing bandwidths of 64\,MHz at 50\,cm
and 1024\,MHz at 10\,cm. Most observations at 20\,cm are made using
the central beam of the Parkes multibeam receiver although the
``H-OH'' receiver has occasionally been used for observations in this
band. Data are simultaneously recorded using the CPSR2 baseband
recording system with 2-bit sampling of two 64-MHz bands, centred on
1341 and 1405\,MHz respectively, and either the WBC or DFB1 with 256
MHz bandwidth. At 50\,cm, data are recorded using one band of
CPSR2. For all receivers, a linearly polarised broad-band calibration
signal can be injected into the feed at $45\deg$ to the two signal
probes.

\begin{table*}
\caption{Parameters for the PPTA pulsars.}
\label{tb:sum}
\begin{center}\begin{tabular}{lrrrr@{.}llr@{}lr@{.}l}
\hline
\multicolumn{1}{c}{PSR} &\multicolumn{1}{c}{$P$} & \multicolumn{1}{c}{$\dot{P}$} & 
\multicolumn{1}{c}{$P_b$} &\multicolumn{2}{c}{DM} &  
\multicolumn{1}{c}{Dist.} & \multicolumn{2}{c}{$\mu$} & 
\multicolumn{2}{c}{$b_E$} \\
 & \multicolumn{1}{c}{(ms)} &\multicolumn{1}{c}{($10^{-20}$)} &\multicolumn{1}{c}{(d)} & 
\multicolumn{2}{c}{($\rm cm^{-3}pc$)} & \multicolumn{1}{c}{(kpc)} & 
\multicolumn{2}{c}{($\rm \mas\ yr^{-1}$)} & \multicolumn{2}{c}{($^{\circ}$)} \\     
\hline
J0437$-$4715 &  5.757 &  5.73 &  5.74 &  2&6   & 0.16$^a$ & 140&.89 & $-67$&87  \\ 
J0613$-$0200 &  3.062 &  0.96 &  1.20 & 38&8   & 1.71 &   7&.3  & $-25$&41  \\
J0711$-$6830 &  5.491 &  1.49 &  ...  & 18&4   & 0.86 &  21&.9  & $-82$&89  \\
J1022$+$1001 & 16.453 &  4.33 &  7.81 & 10&3  & 0.30$^a$ &  17&    & $ -0$&064 \\
J1024$-$0719 &  5.162 &  1.85 &  ...  &  6&5    & 0.39 &  81&    & $-16$&04  \\
J1045$-$4509 &  7.474 &  1.77 &  4.08 & 58&1    & 1.96 &   7&.8  & $-47$&71  \\
J1600$-$3053 &  3.598 &  0.95 & 14.35 & 52&3    & 1.63 &   4&.1  & $-10$&07  \\
J1603$-$7202 & 14.842 &  1.56 &  6.31 & 38&1   & 1.17 &   8&.5  & $-49$&96  \\
J1643$-$1224 &  4.622 &  1.85 &147.02 & 62&4   & 2.41 &   9&    & $  9$&78  \\
J1713$+$0747 &  4.570 &  0.85 & 67.83 & 16&0   & 1.12$^a$ &   6&.4  & $ 30$&70  \\
J1730$-$2304 &  8.123 &  2.02 &  ...  &  9&6    & 0.53 &  20  &.5    & $  0$&19  \\
J1732$-$5049 &  5.313 &  1.38 &  5.26 & 56&8    & 1.41$^a$ &  --  & --    & $-27$&49  \\
J1744$-$1134 &  4.075 &  0.89 &  ...  &  3&1  & 0.36 &  20&.99 & $ 11$&81  \\
J1824$-$2452 &  3.054 &162.00 &  ...  &119&9    & 3.09 &   4&.7  & $ -1$&55  \\
J1857$+$0943 &  5.362 &  1.78 & 12.33 & 13&3    & 0.91$^a$ &   6&.16 & $ 32$&32  \\
J1909$-$3744 &  2.947 &  1.40 &  1.53 & 10&4  & 1.14$^a$ &  36&.99 & $-15$&16  \\
J1939$+$2134 &  1.558 & 10.50 &  ...  & 71&0   & 3.57 &   0&.80 & $ 42$&30  \\
J2124$-$3358 &  4.931 &  2.05 &  ...  &  4&6   & 0.27 &  49&.0  & $-17$&82  \\
J2129$-$5721 &  3.726 &  2.07 &  6.63 & 31&9    & 1.36 &   8&    & $-39$&90  \\
J2145$-$0750 & 16.052 &  2.98 &  6.84 &  9&0   & 0.50$^a$ &  14&.1  & $  5$&31 \\ 
\hline
\end{tabular}\end{center}
$^a$ Distance obtained from a parallax measurement.
\end{table*}

Observation times per pulsar are typically either 32\,min or 64\,min
and data are folded on-line with sub-integration times of 1\,min for
the WBC and DFB1 and 16\,s for CPSR2. All pulsar observations are
preceded by a short (2\,min) observation of a pulsed calibration
signal. For most pulsars the WBC and DFB1 data are split into 512
frequency channels with between 256 and 1024 phase bins across the
pulse period. For CPSR2, the data are coherently dedispersed in each
of 128 frequency channels with 1024 phase bins. Off-line processing
uses the \textsc{psrchive} software system \citep{hvm04}. For all
recording systems, data from frequency channels or sub-integrations
which are obviously affected by radio-frequency interference are
excised, as are channels from the outer edges of the band (typically
about 5 per cent of the band at each edge) where the system gain is
low. Data are then calibrated for variations of instrumental gain and
phase across the band using observations of the pulsed calibration
signal and the Stokes parameters formed.

\begin{figure*}
 \includegraphics[width=12cm,angle=-90]{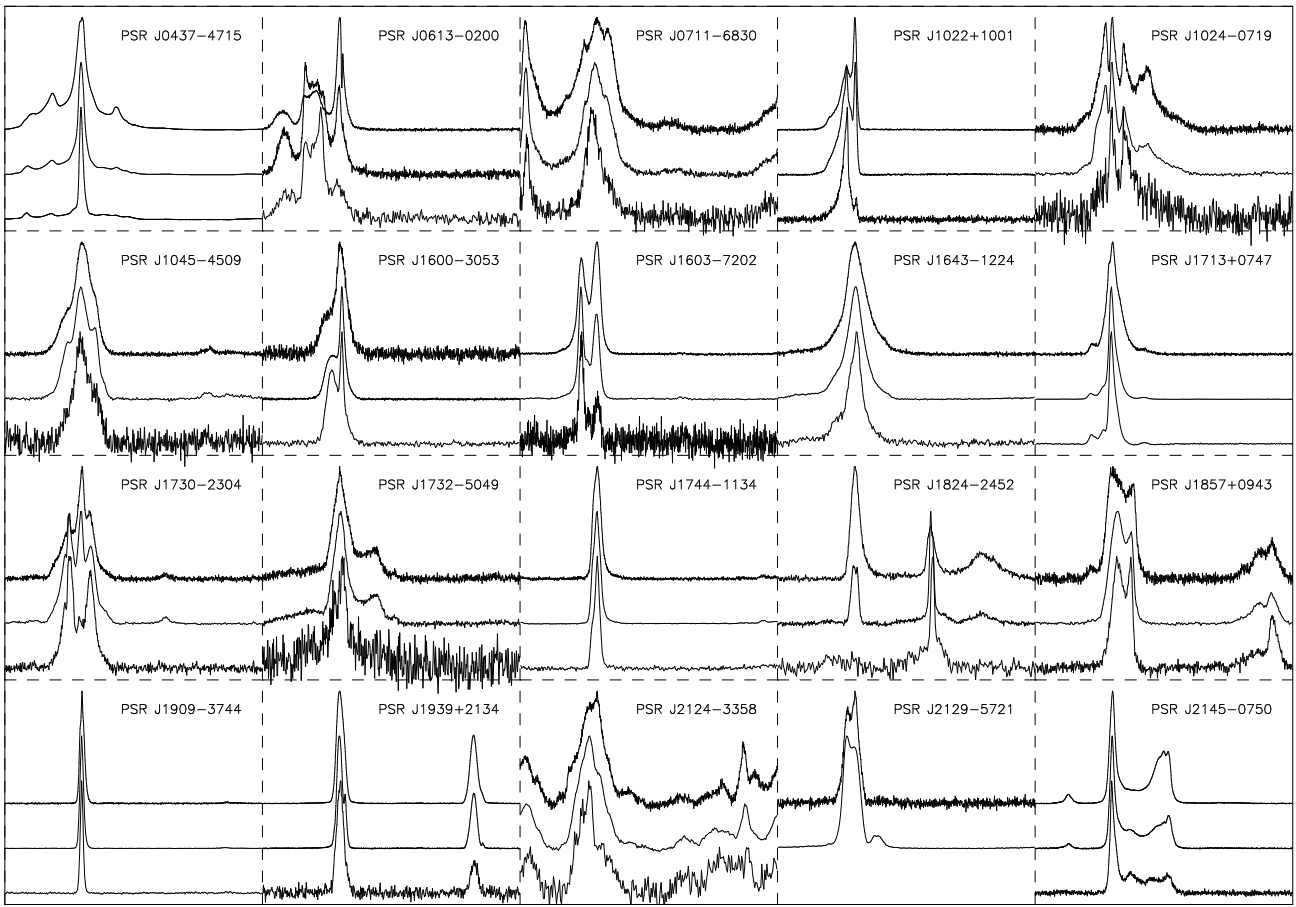}
 \caption{Stokes I profiles for the 20 millisecond pulsars in our
 sample.  For each pulsar we give the 50\,cm, 20\,cm and 10\,cm
 profiles from top to bottom respectively (except for PSR~J2129$-$5721
 where only 50\,cm and 20\,cm profiles are available).}\label{fg:profiles}
\end{figure*}

For all pulsars except PSR~J0437$-$4715, pulse TOAs were obtained by
cross-correlating a template profile with the Stokes I mean pulse
profile for each observation. For most of the observed pulsars, errors
in the calibration procedure resulted in TOA errors which were less
than the uncertainty due to random (receiver) noise. However for
PSR~J0437$-$4715 at 20\,cm and 50\,cm, this was not the case and it
was advantageous to use the polarimetric invariant interval instead of
Stokes I \citep{bvb+00}. Template profiles were formed for each
instrument and each observing frequency (685, 1341, 1405\,MHz for
CPSR2, 1369 and 3100\,MHz for the DFB1, 1433 and 3100\,MHz for the
WBC) by weighted summing of all available data to form `grand average'
profiles and then blanking the baseline
regions. Figure~\ref{fg:profiles} shows the grand average Stokes I
profiles at the three frequencies for all 20 pulsars (except for
PSR~J2129$-$5721 where we have data for two frequencies only).

For DM measurements, profile alignment across frequencies is an
important issue. The template profiles for a given pulsar were
approximately aligned using the cross-correlation of each profile with
a reference profile. However, because of frequency-dependent profile
variations, there remains some uncertainty in the true alignment. In
this work, we are primarily concerned with variations in DM, not
absolute values, so arbitrary phase offsets between data obtained
using different systems were included in the timing model.

The resulting TOAs were analysed using {\sc tempo2}. Timing model
parameters were obtained by fitting standard pulsar timing parameters
(including astrometric, spin and binary parameters) to the 20\,cm
and/or the 10\,cm observations\footnote{We use residuals obtained
using CPSR2, WBC and DFB1. However, for a few pulsars with high DM and
short period, the WBC profile is significantly smeared and therefore,
for PSRs~J0613$-$0200, J1600$-$3053, J1824$-$2452, J1939$+$2134, we
only use the CPSR2 and DFB1 observations.}. As we are interested in DM
variations, we do not fit for any time derivatives of the DM as part
of the timing model; however, we do allow \textsc{tempo2} to model the
DM variations due to the solar wind (this is further discussed in
\S\ref{sec:swind}). The timing model parameters were subsequently held
constant in order to obtain timing residuals at all observing
frequencies.

To obtain the time variations in DM, $\Delta \mbox{DM}(t)$, we fitted
Equation~\ref{eq:dm1} to segments of timing data. The segment lengths
were adjusted so that each segment contained at least one observation
at each frequency (typically 1 or 2 weeks).  Initially we used the 10
and 50\,cm observations to obtain $\Delta \mbox{DM}(t)$ because these
were obtained simultaneously and are well separated in frequency.
However, if the pulse profile at 10\,cm or 50\,cm has a low
signal-to-noise ratio, we also used 20 and 50\,cm or 10 and 20\,cm to
obtain $\Delta \mbox{DM}(t)$.

The structure function is a useful statistic for studying the physical
process causing the DM variations. Calculation of the structure
function is straightforward, even for unequally-spaced data
(Equation~\ref{eq:stdm}). However, the estimation of the errors in the
structure function due to uncertainties in the $\Delta \mbox{DM}(t)$
values, and due to the finite duration of the $\Delta \mbox{DM}(t)$
series has not been discussed consistently in the literature. Because
of the irregular data-sampling, $\tau$ represents a ``bin'' with a
width which we have adjusted to give roughly equal logarithmic
sampling.  Estimation of $D_{\rm DM}(\tau)$ for a power-law process from a
single ``realisation'' of the process incurs a significant error which
has been discussed by \citet{rcm00}.  For Kolmogorov processes they
found that the estimation error $\sigma_{\rm est}(\tau) \propto
D(\tau)[\tau/(T-\tau)]^{1/3}$ where $T$ represents the data-span.  We
have extended their simulations to pure power-law processes which do
not have a defined low frequency limit (or ``outer scale'').  For
uniformly sampled data from a Kolmogorov process we find that
\begin{equation}
\sigma_{\rm est}(\tau) =1.66D(\tau)(\tau/T)^{1/3}.
\end{equation}
Assuming that the process itself has Gaussian differences, the
structure function estimator must be $\chi^2$-distributed, and defined
by $N_{\rm dof}$, the number of degrees of freedom. The estimation
error can be written in terms of $N_{\rm dof}$ as $\sigma_{\rm
est}(\tau) = D(\tau)(2/N_{\rm dof})^{0.5}$, so $N_{\rm
dof}=0.72(T/\tau)^{2/3}$. As our data are irregularly sampled we
sometimes have fewer pairs, $N_p$, or fewer samples, $N_s$,
contributing than $N_{\rm dof}$. We therefore approximate the actual
number of degrees of freedom as the minimum of $(N_p,N_s,N_{\rm
dof})$.

In addition to the estimation error discussed above, which is the
amount by which a single realisation of a random process can be
expected to depart from the theoretical mean, we must consider the
fact that the measurements include a white noise component independent
of ${\rm DM}(t)$. We assume that the errors on the $\Delta
\mbox{DM}(t)$ values are independent, Gaussian and have known, but
different, standard deviations.  These contribute a different bias and
error to each $D_{\rm est}(\tau)$ which are computed by expanding each
$D_{\rm est}(\tau)$ as shown in Appendix A.

Our work contrasts with earlier estimates of the structure function.
In earlier work, the uncertainty on the structure function was either
estimated as the standard deviation divided by the square root of the
number of points \citep{cwd+90} or as the smaller of the number of
points and $(T/\tau)$ \citep{rdb+06}. The choice of the bias term
which is subtracted from $D_{\rm DM}(\tau)$ has also been inconsistent in
the literature. Usually the bias has been taken as the average of the
$\Delta \mbox{DM}(t)$ errors. This method is only accurate when the
number of points is very large and the error on $\Delta \mbox{DM}(t)$
is significantly smaller than its value.  Earlier work also did not
allow for the error on the measured $\Delta \mbox{DM}(t)$ values
which, for some data-sets, is very important.

It is useful to obtain the diffractive time-scale, $\tau_d$, and
bandwidth, $\nu_d$, for each pulsar to compare with the
$D_{\rm DM}(\tau)$. For many pulsars we were able to obtain these values
from the literature. Table~\ref{tb:taud} gives these $\tau_d$ and
$\nu_d$ values. In column order, the Table contains the pulsar name,
observing frequency, $\tau_d$, $\nu_d$ and a bibliographic
reference. However, no measurements existed for nine of our
pulsars. As our data have relatively poor frequency and time
resolution (for this purpose), we obtained $\tau_d$ and $\nu_d$ using
a structure function analysis \citep{rcm00} rather than the more
standard method of using the auto-correlation function of the dynamic
spectrum. The implementation of this technique is discussed in
Appendix B.  From each $\tau_d$ measurement we have used the
definition that $D_\phi(\tau_d) = 1$ to obtain an estimate of
\begin{eqnarray}
D_{\rm DM}(\tau_d)=(K\nu/2\pi)^{2}.
\end{eqnarray}
  
\begin{table}
  \caption{Scintillation parameters for the PPTA millisecond pulsars}
\label{tb:taud}
\setlength{\tabcolsep}{1mm}
\begin{center}
\begin{tabular}{lrcclrrrrrrl}
\noalign{\smallskip}
\hline
PSR Name & \multicolumn{1}{c}{Freq.} &
\multicolumn{1}{c}{$\tau_d$}  &\multicolumn{1}{c}{$\nu_d$}&  Ref.\\
       & \multicolumn{1}{c}{(MHz)} & \multicolumn{1}{c}{(min)} & \multicolumn{1}{c}{(MHz)} &   \\
  \hline
J0437$-$4715 &    327   & 1.9 -- 5.1 & 0.18 -- 3.0  & 1 \\
     $ $     &    436   & 4.6 -- 11  & 3.2 -- 4.4  &2,3 \\
     $ $     &    660   & 7.8        & 17          &2 \\
J0711$-$6830 &    436   & 13         & 0.37        &    2 \\
     $ $     &    660   & 16         & 1.2         &    2 \\
J1600$-$3053 &   1373   &  4.7       & $<$ 0.5     &    4 \\
J1603$-$7202 &    660   &  9.2       & 0.36        &    2 \\
J1713$+$0747 &    430   & 14         & 0.6         &    5 \\
     $ $     &    436   & 28         & 1.5         &    2 \\
J1730$-$2304 &    327   & 7.4 -- 7.5 &0.10 -- 0.12 &    1 \\
     $ $     &    436   & 6.3 -- 12  &0.15 -- 0.18 &    2,3 \\
     $ $     &    660   & 9.7        &1.4          &    2 \\
     $ $     &   1520   & 23-27      & 30 -- 38    &    2,3 \\
J1744$-$1134 &    436   & 21         & 1.3         &    2 \\
     $ $     &    660   & 20         & 2.3         &    2 \\
J1939$+$2134 &    320   &  1.1       & 0.0014      &    6 \\
     $ $     &    430   &  1.7       & 0.0042      &    6 \\
     $ $     &   1400   &  7.4       & 0.92        &    6 \\
J2124$-$3358 &    436   & 44         & 6.9         &    2 \\
J2129$-$5718 &    436   & 11         & 0.29        &    2 \\
     $ $     &    660   & 17         & 1.3         &    2 \\
     $ $     &   1520   & 24         & 58          &    2 \\
J2145$-$0750 &    327   &  6.4       & 0.33        &    1 \\
             &    436   & 21-25      & 0.61 -- 2.5 &    2,3 \\
  \hline
\end{tabular}
\end{center}
Reference:
(1) \citet{gg00}; (2) \citet{jnk98}; (3) \citet{nj95};
(4) \citet{ojhb06}; (5) \citet{bplw02}; (6) \citet{cwd+90}.

\end{table}

\section{Results}

\begin{figure*}
  \centering
  \begin{tabular}{c}
    \includegraphics[width=170mm]{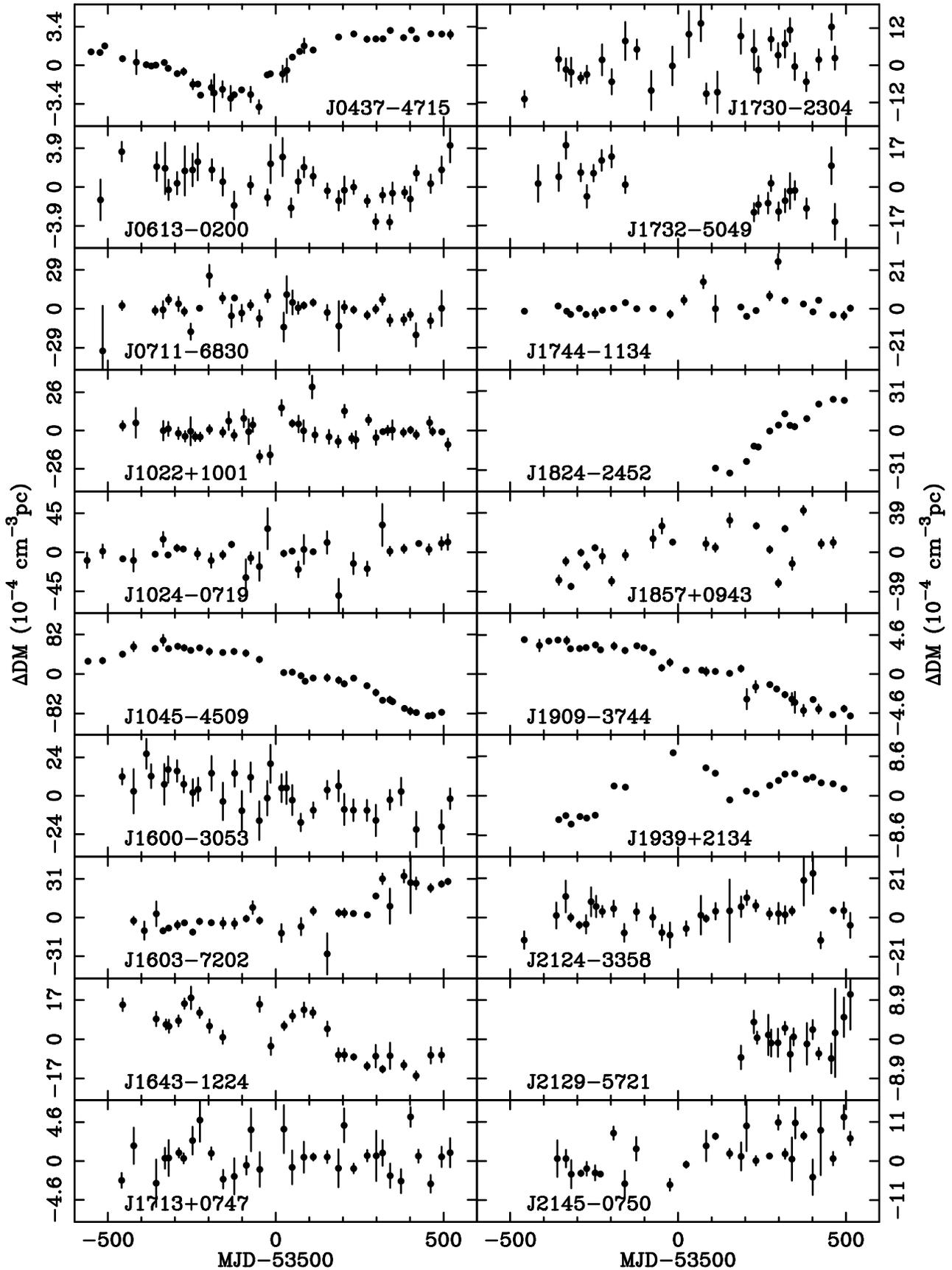}
  \end{tabular}
  \caption{DM variations of 20 millisecond pulsars. Note a $\Delta$DM
  of $10^{-4}\rm cm^{-3}pc$ corresponds to time delays at 10\,cm of
  43\,ns, at 20\,cm of 212\,ns and at 50\,cm of 884\,ns.}
  \label{fg:DMvar}
\end{figure*}

The variation of DM with time, $\Delta \mbox{DM}(t)$, for each pulsar
is shown in Figure~\ref{fg:DMvar}.  We list,
in Table~\ref{tb:DM}, the bands used for each pulsar and the interval
over which the $\Delta \mbox{DM}(t)$ values were measured and the
slope of the best-fitting straight line ${\rm dDM/d}t$. The panels in
Figure~\ref{fg:DMvar} are chosen so that all
pulsars have the same time axis, but different scalings are used for
the ordinate.  We see large-scale DM variations for six of our pulsars
(PSRs~J0437$-$4715, J1045$-$4509, J1643$-$1224, J1824$-$2452,
J1909$-$3744, J1939$+$2134) with a maximum range in $\Delta$DM of
$\sim 0.014$\,cm$^{-3}$\,pc for PSR~J1045$-$4509.

\begin{table}
\caption{Summary of the DM variations for the PPTA pulsars}
\label{tb:DM}
\begin{center}\begin{tabular}{ccrrc}
\hline
 
\multicolumn{1}{c}{PSR~Name} & \multicolumn{1}{c}{Band}& Interval &
\multicolumn{1}{c}{$\mbox {dDM}/ \mbox{d}t $} &
\multicolumn{1}{c}{Data Span} \\
 & (cm) & (d) & ($\rm cm^{-3}\ pc\ yr^{-1}$) & (yr)   \\ 

\hline
J0437$-$4715 & 10, 50 & 15 & $ 1.0(2) \times10^{-5}$ & 3.0 \\
J0613$-$0200 & 20, 50 & 15 & $-9(2)   \times10^{-5}$ & 2.9 \\
J0711$-$6830 & 20, 50 & 15 & $-2.5(9) \times10^{-5}$ & 2.8 \\
J1022$+$1001 & 10, 50 &  7 & $-5(60)  \times10^{-6}$ & 2.7 \\
J1024$-$0719 & 20, 50 & 15 & $ 3.2(9) \times10^{-4}$ & 2.9 \\
J1045$-$4509 & 20, 50 & 15 & $-5.56(9)\times10^{-3}$ & 2.9 \\
J1600$-$3053 & 10, 20 & 15 & $-9(2)   \times10^{-4}$ & 2.7 \\
J1603$-$7202 & 20, 50 & 15 & $1.28(5) \times10^{-3}$ & 2.6 \\
J1643$-$1224 & 20, 50 & 15 & $-1.18(6)\times10^{-3}$ & 2.6 \\
J1713$+$0747 & 20, 50 & 15 & $5(23)   \times10^{-6}$ & 2.7 \\
J1730$-$2304 & 20, 50 &  7 & $3.9(8)  \times10^{-4}$ & 2.5 \\
J1732$-$5049 & 20, 50 &  7 & $-7(1)   \times10^{-4}$ & 2.4 \\
J1744$-$1134 & 20, 50 & 15 & $ 5(2)   \times10^{-5}$ & 2.7 \\
J1824$-$2452 & 20, 50 & 12 & $ 6.0(1) \times10^{-3}$ & 1.1 \\
J1857$+$0943 & 20, 50 & 15 & $ 1.5(1) \times10^{-3}$ & 2.2 \\
J1909$-$3744 & 20, 50 & 15 & $-3.28(6)\times10^{-4}$ & 2.7 \\
J1939$+$2134 & 20, 50 & 15 & $ 2.57(2)\times10^{-4}$ & 2.3 \\
J2124$-$3358 & 20, 50 & 15 & $ 2.5(8) \times10^{-4}$ & 2.7 \\
J2129$-$5721 & 20, 50 &  7 & $-2(3)   \times10^{-4}$ & 0.9 \\
J2145$-$0750 & 20, 50 & 15 & $ 4.0(4) \times10^{-4}$ & 2.4 \\
\hline
\end{tabular}\end{center}
\end{table}

\subsection{Diffractive scintillation parameters}

Using the method described in \S3, we obtained diffractive
scintillation timescales for 17 of our pulsars, obtaining $\tau_d$ and
$\nu_d$ values for each observation with a high S/N.  In column order,
Table~\ref{tb:ourtaud} contains the pulsar name, observing frequency
and the range of our measured $\tau_d$ and $\nu_d$ values,
respectively.  For pulsars where previous measurements exist our
results are consistent with values in the literature. The scatter in
$\tau_d$ observations is much greater than the error bars on
individual $\tau_d$ measurements. We have confirmed, by simulation,
that the reason for this is that $\tau_d$ is estimated from
observations which are much shorter than the refractive scale.

For three pulsars it was difficult for us to obtain diffractive
time-scale measurements. For PSR~J0437$-$4715, this is not a problem
as there are many measurements available in the literature. The
diffractive time-scale for PSR~J1824$-$2452 is too short at 20\,cm
(less than 1\,min) for us to measure.  At 10\,cm, this pulsar is very
weak (SNR $\sim$20 for a 1\,hr observation), but we were able to
obtain a few usable observations.  The $\tau_d$ for PSR~J2124$-$3358
is relatively long. From the $\tau_d=44$\,min at 436\,MHz
\citep{jnk98}, we can estimate that $\tau_d$ at 685\,MHz is
$\sim$76\,min, but our current observation time for this pulsar is
only 32\,min.

\begin{table}
  \caption{Scintillation parameters from our observations}
\label{tb:ourtaud}
\setlength{\tabcolsep}{1mm}
\begin{center}
\begin{tabular}{lrr@{ $-$ }lr@{ $-$ }l}
\noalign{\smallskip}
\hline
PSR Name & \multicolumn{1}{c}{Freq.} &\multicolumn{2}{c}{$\tau_d$} &\multicolumn{2}{c}{$\nu_d$}  \\
         & \multicolumn{1}{c}{(MHz)} &\multicolumn{2}{c}{(min)} &\multicolumn{2}{c}{(MHz)} \\
  \hline
J0613$-$0200 &    1369  &10   & 54  & 0.98 & 3.1\\
J0711$-$6830 &    685   &18   & 42  & 1.0  & 5.4 \\
             &    1369  &36   & 127 & 30   & 77 \\
J1022$+$1001 &    685   &53   & 184 & 6.4  & 40 \\
J1024$-$0719 &    685   &23   & 89  & 4.8  & 47 \\
J1045$-$4509 &   3100   &2.2  & 12  & 0.64 & 15\\
J1600$-$3053 &   3100   &4.0  & 23  & 1.90 & 6.9 \\
J1603$-$7202 &    685   &8.6  & 27  & 1.5  & 7.8\\
             &   1369   &7.7  & 40  & 1.8  & 18\\
J1643$-$1224 &   3100   &2.6  & 12  & 0.89 & 2.0\\
J1713$+$0747 &    685   &20   & 47  & 1.8  & 11 \\
J1730$-$2304 &    685   &9.3  & 28  & 1.2  & 5.4\\
     $ $     &   1369   &17   & 54  & 3.9  & 32 \\
J1732$-$5049 &    1369  &18   & 39  & 1.8  & 6.2\\
J1744$-$1134 &    685   &29   & 76  & 4.1  & 40 \\
J1824$-$2452 &   3100   &1.1  & 9.5 & 0.6  & 1.1    \\
J1857$+$0943 &    685   &13   & 22  & 2.5  & 7.8\\
             &   1369   &16   & 68  & 2.7  & 25\\
J1909$-$3744 &    685   &16   & 69  & 2.8  & 20 \\
J1939$+$2134 &   1369   &4.1  & 10  & 1.8  & 5.4\\
J2129$-$5718 &    685   &12   & 39  & 1.7  & 4.5\\
     $ $     &   1369   &35   & 79  & 25  & 234\\
J2145$-$0750 &    685   &20   & 111 & 2.9  & 44\\
\hline
\end{tabular}
\end{center}
\end{table}

\subsection{Structure functions}

\begin{figure*}
  \centering
    \includegraphics[angle=-90, width=90mm]{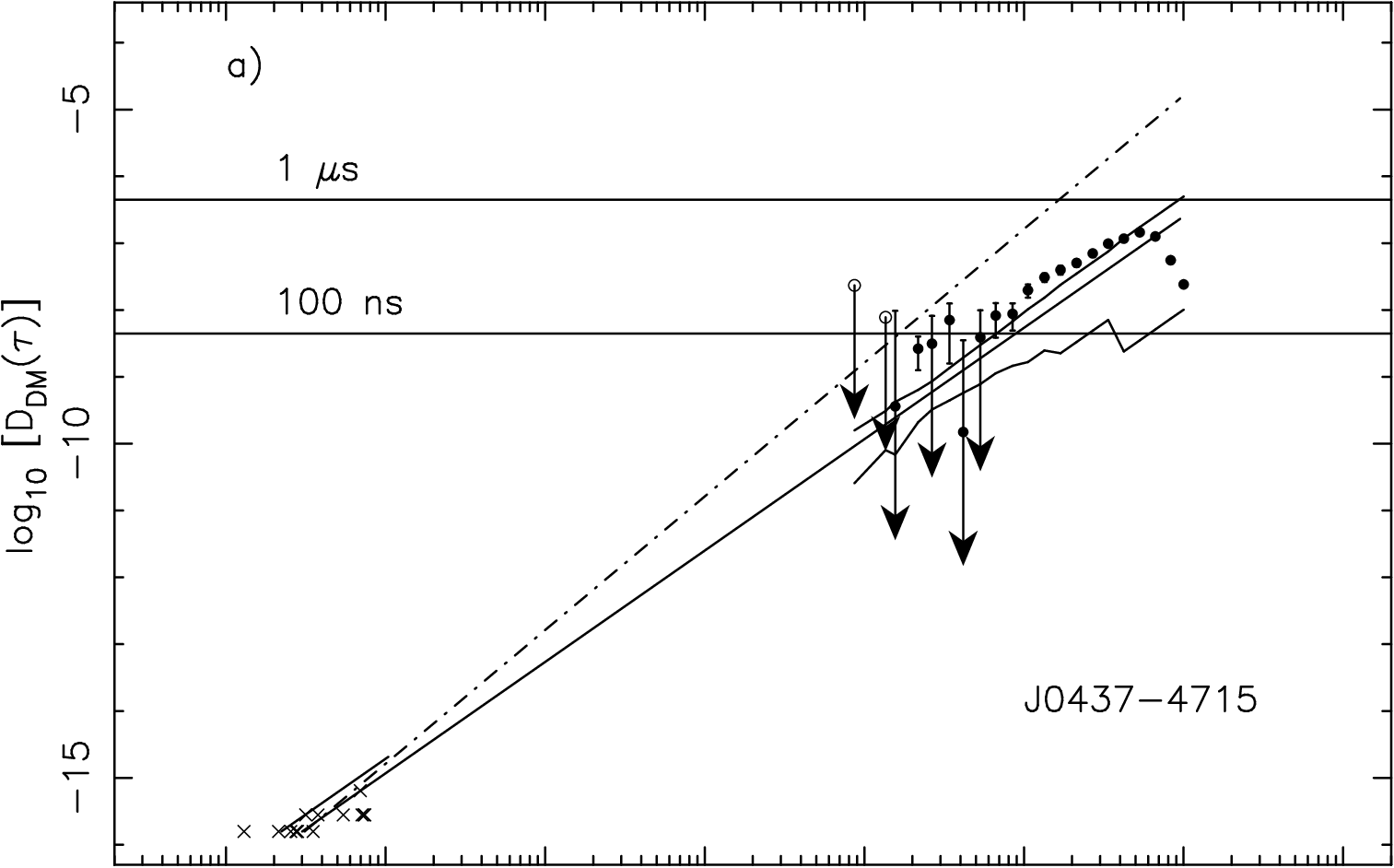}
    \includegraphics[angle=-90, width=82.8mm]{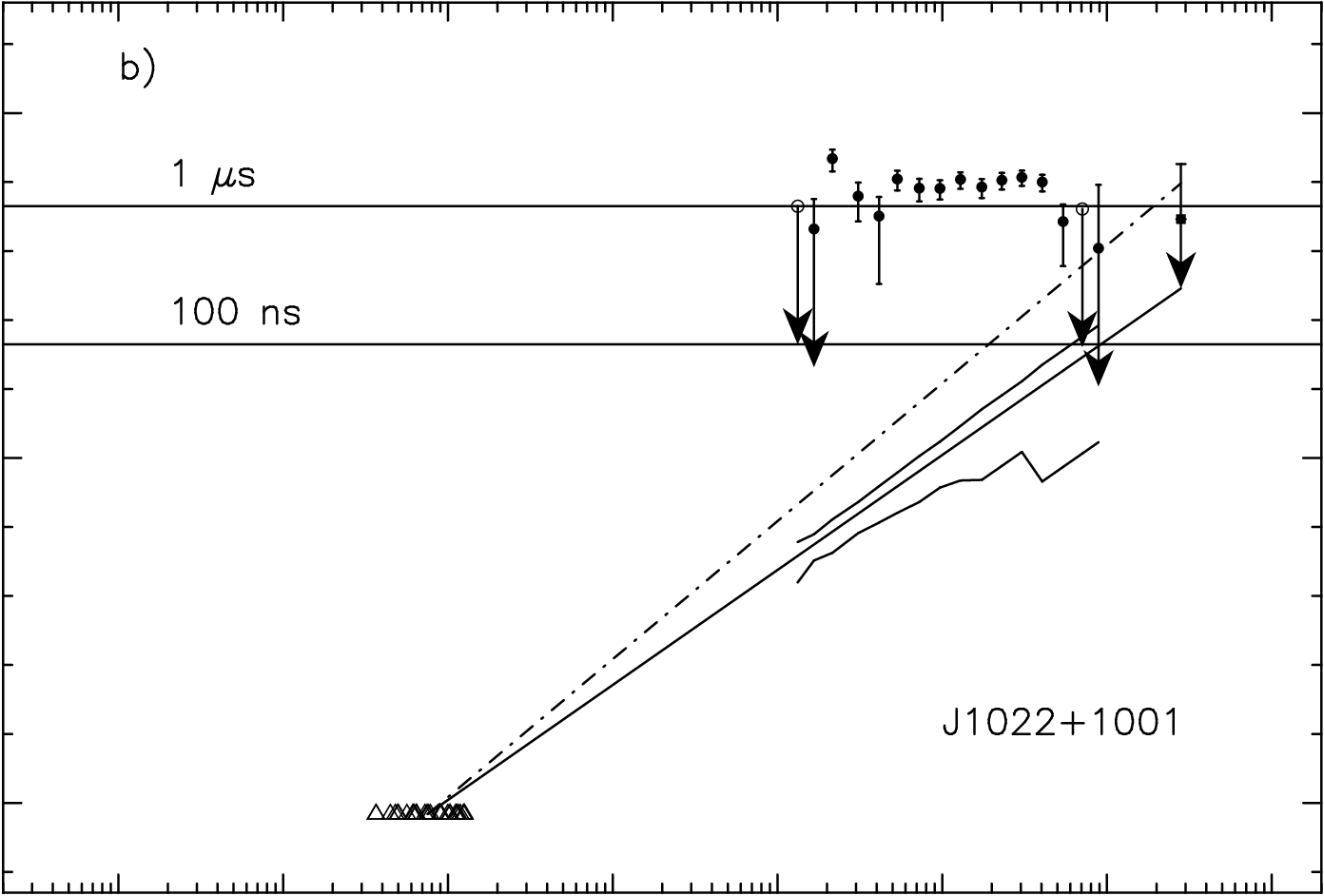}\\
    \includegraphics[angle=-90, width=90mm]{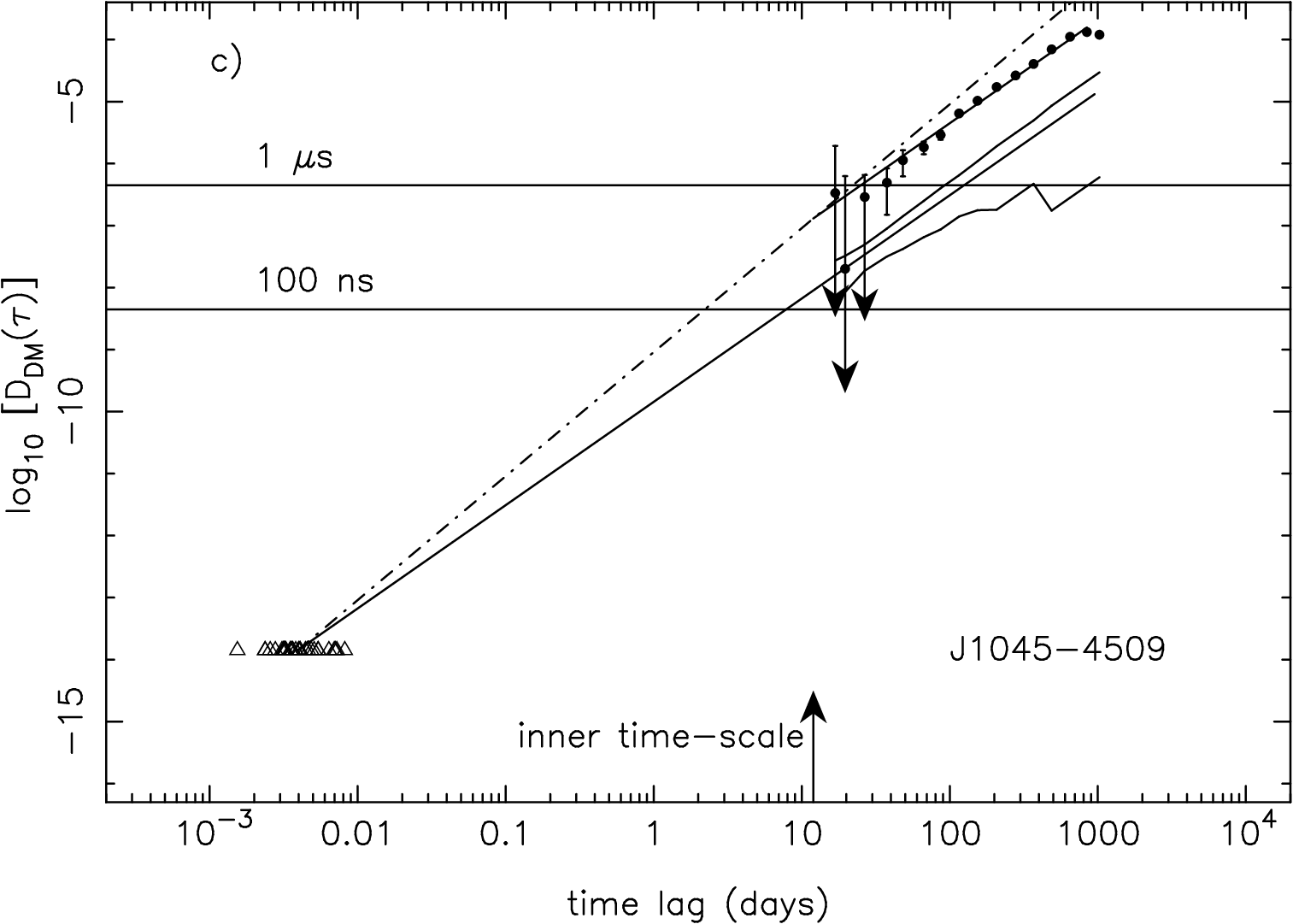}
    \includegraphics[angle=-90, width=82.8mm]{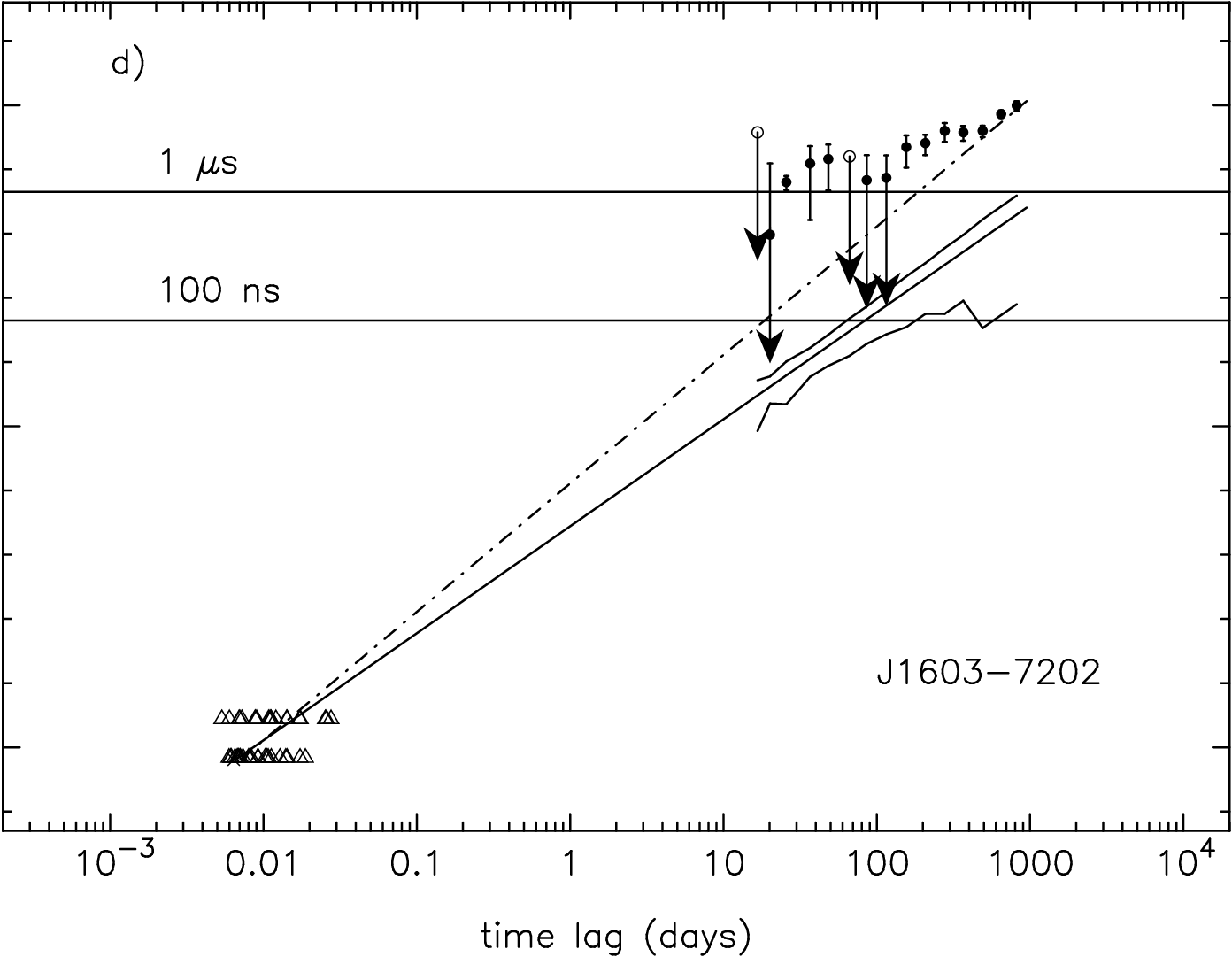}
  \caption{Structure functions, $D_{\rm DM}(\tau)$ for four
  pulsars. The $\tau_d$ derived estimates are marked by triangles from
  our data and crosses from the literature.  The estimates obtained
  directly from the time series $\Delta \mbox{DM}(t)$ as discussed in
  Appendix A are marked as filled circles with error bars. Open
  circles indicate a negative estimate.  A Kolmogorov model fit to
  $\tau_d$ is shown using a solid line and a quadratic model is shown
  dash-dotted.  Confidence limits on the Kolmogorov model are solid
  lines bracketting the model.  Equivalent delays at 1400\,MHz are
  shown for 100\,ns and 1\,$\mu s$. For PSR~J1022$+$1001, a point
  derived from $\mbox {dDM}/ \mbox{d}t $ is shown as a solid box with
  error bars at the longest time lag.}
  \label{fg:sf1}

\end{figure*}

\begin{figure*}
  \centering
    \includegraphics[angle=-90,width=90mm]{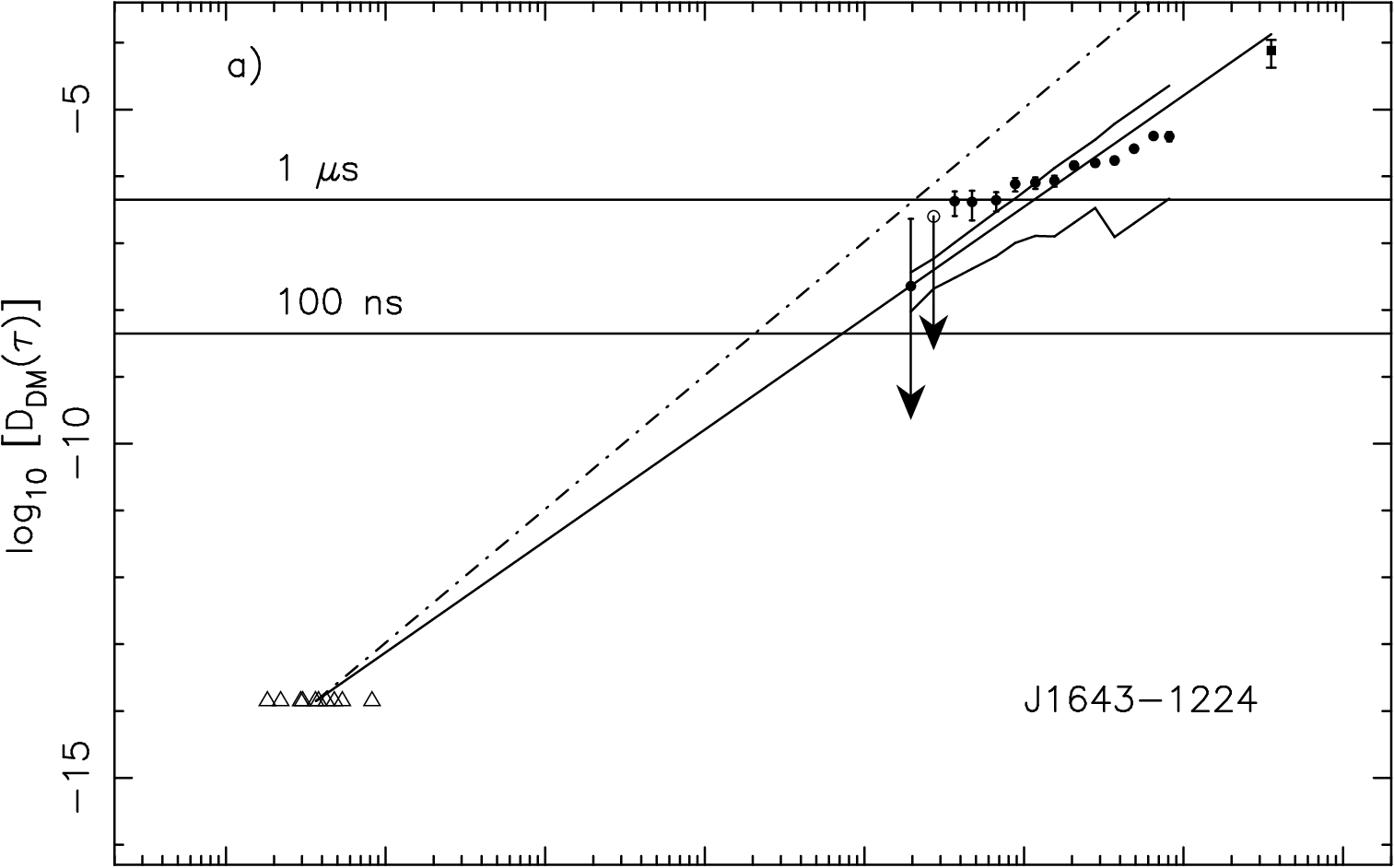}
    \includegraphics[angle=-90,width=82.8mm]{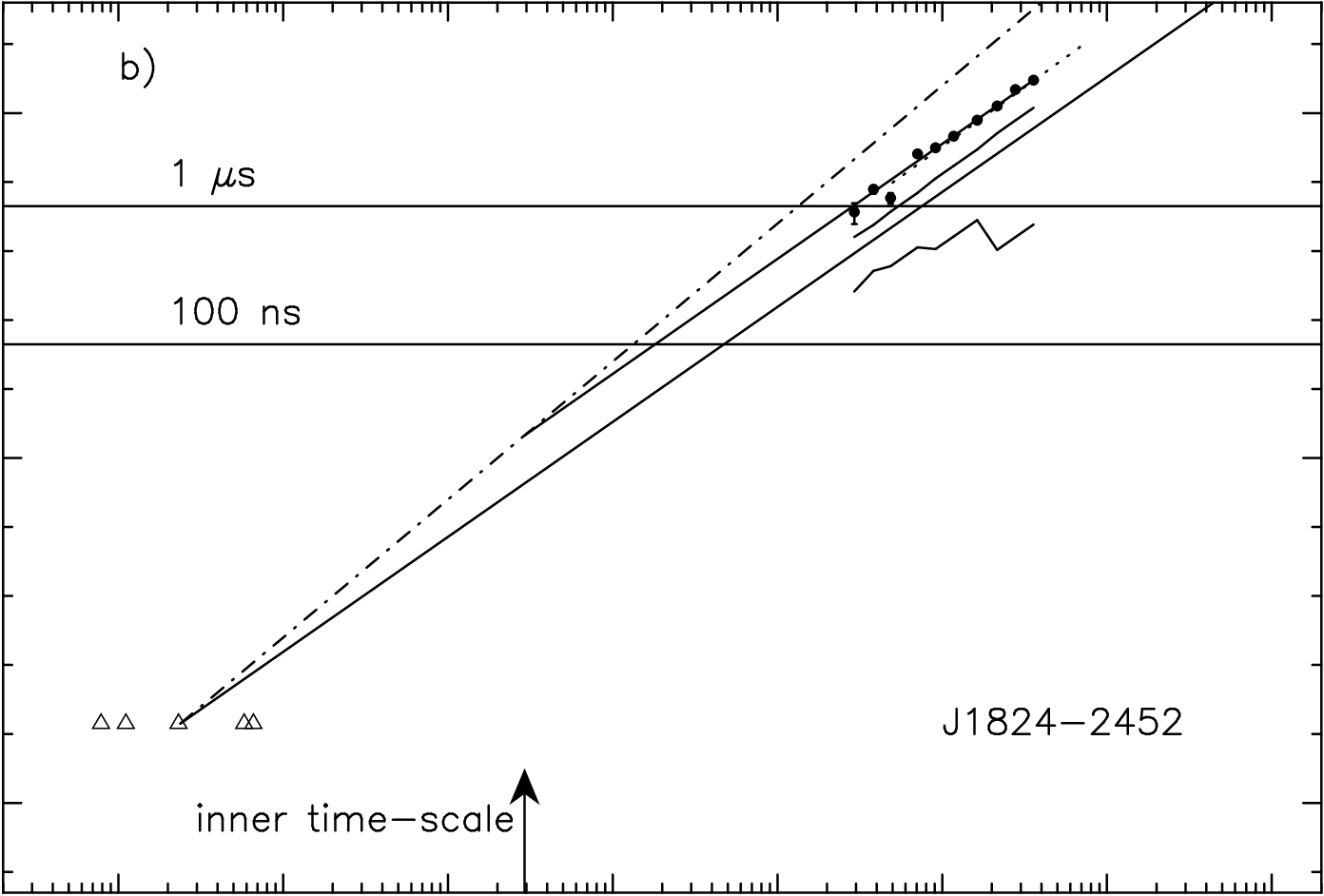}\\
    \includegraphics[angle=-90, width=90mm]{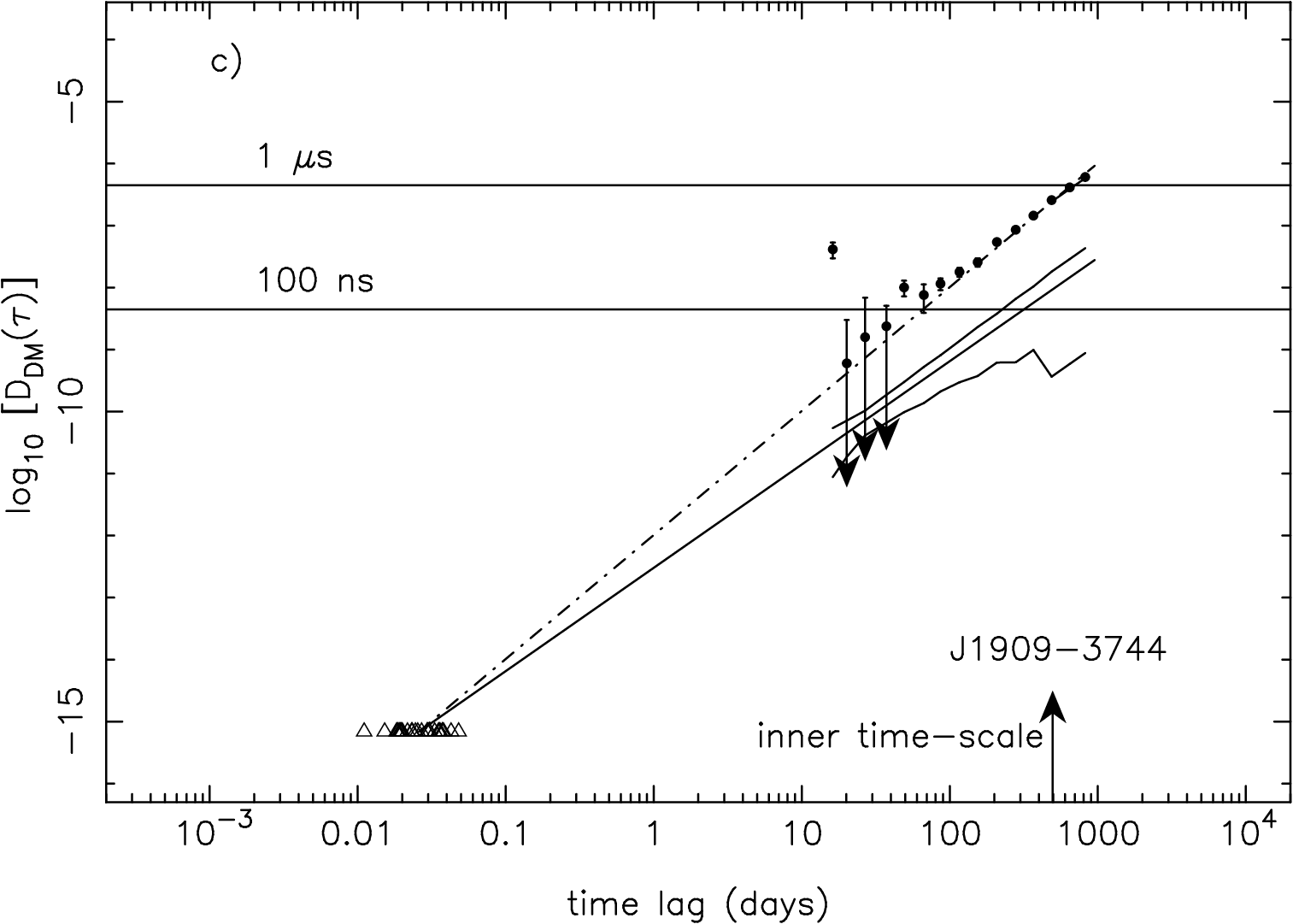}
    \includegraphics[angle=-90, width=82.8mm]{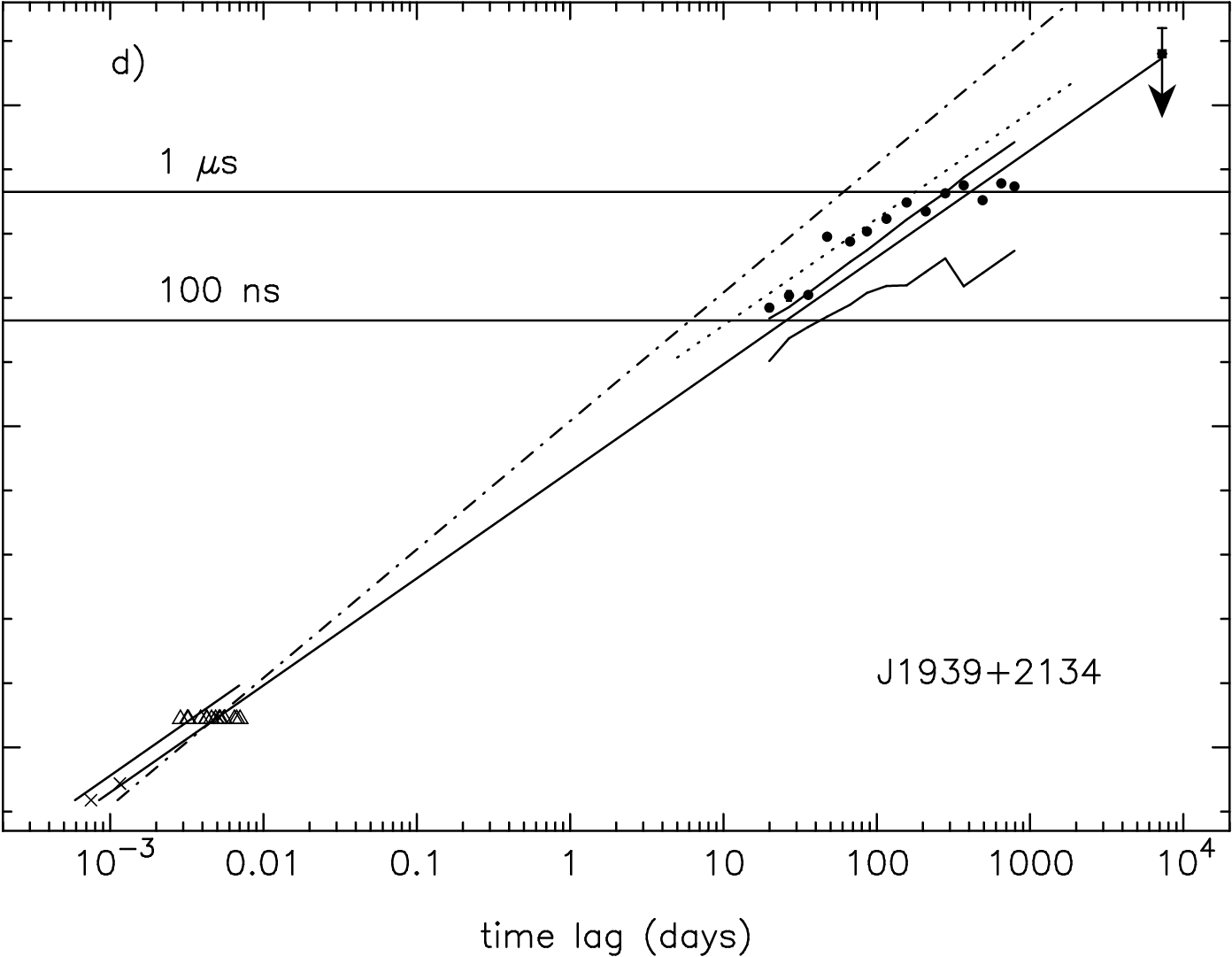}
  \caption{As for Figure~\ref{fg:sf1}, but for another four pulsars. For
  PSRs~J1643$-$1224 and J1939$+$2134, a point derived
  from $\mbox {dDM}/ \mbox{d}t $ is shown as a solid box with error
  bars at the longest time lag.}
  \label{fg:sf2}

\end{figure*}

We have calculated structure functions from $\Delta \mbox{DM}(t)$ for
each of our pulsars.  Representative examples are shown in
Figure~\ref{fg:sf1} and Figure~\ref{fg:sf2}. In these figures, we have
included $\tau_d$ measurements obtained from the literature
(cross-symbols) or from our data (open triangle symbols).  For some
pulsars, we have been able to derive an estimate of $D_{\rm DM}$ at
large time lags from ${\rm dDM/d}t$ measurements in the literature
\citep{hlk+04} that were obtained using a single data-set\footnote{ A
given $\mbox {dDM}/ \mbox{d}t $ measurement can be converted to a
single point on a structure function as $(\mbox {dDM}/ \mbox{d}t\cdot
T)^2$, where $T$ is the data span.}; such points are indicated using a
full square-symbol at the rightmost edge of the plot. To put the data
in context, we have drawn two theoretical lines fitted through the
$\tau_d$ points, one (full line) with the Kolmogorov exponent
($\alpha=5/3$), the other (dashed-dotted line) with $\alpha=2$
corresponding to the steepest possible structure function resulting
from plasma turbulence \citep{ric90} (hereafter, this spectrum is
known as ``quadratic''). Estimation error bounds on the theoretical
Kolmogorov model (at the 68\% confidence level) for each data point
are plotted using solid lines that bracket the theoretical model.

The remaining symbols used on the figures are as follows. The
structure function values measured from $\Delta \mbox{DM}(t)$ are
plotted using solid circle symbols. The errors on these points are
estimated from the uncertainty on $\Delta \mbox{DM}(t)$.  For cases
where the error is larger than the value we use downward pointing
arrow symbols for the lower bound on the error bar. As we subtracted
the bias due to the uncertainties on the $\Delta \mbox{DM}(t)$
measurements, it is possible for large uncertainties on $\Delta
\mbox{DM}(t)$ that the structure function values are negative. We
indicate such points using open circles and a downward arrow plotted
at $D_{\rm DM}(\tau)$ plus twice its error. The structure function plots
all have the same scaling.

For comparison, we also indicate the value of $D_{\rm DM}$ that would be
expected for white timing residuals with a given rms ($\sigma_{rms}$)
at 1400\,MHz.  The relationship between the structure function of the
timing residuals, $D_{\rm TOA}$, and the structure function of DM
variations, $D_{\rm DM}$, is
\begin{eqnarray}
  D_{\rm DM} = D_{\rm TOA}(K\nu^2)^2.
\label{eq:dtoa}
\end{eqnarray}
If the timing residuals are white, then $D_{TOA}=2\sigma_{rms}^2$.  We
indicate, using solid horizontal lines, white noise with an rms of
1\,$\mu s$ and 100\,ns. 

\begin{figure*}
  \centering
  \begin{tabular}{c}
    \includegraphics[angle=-90, width=90mm]{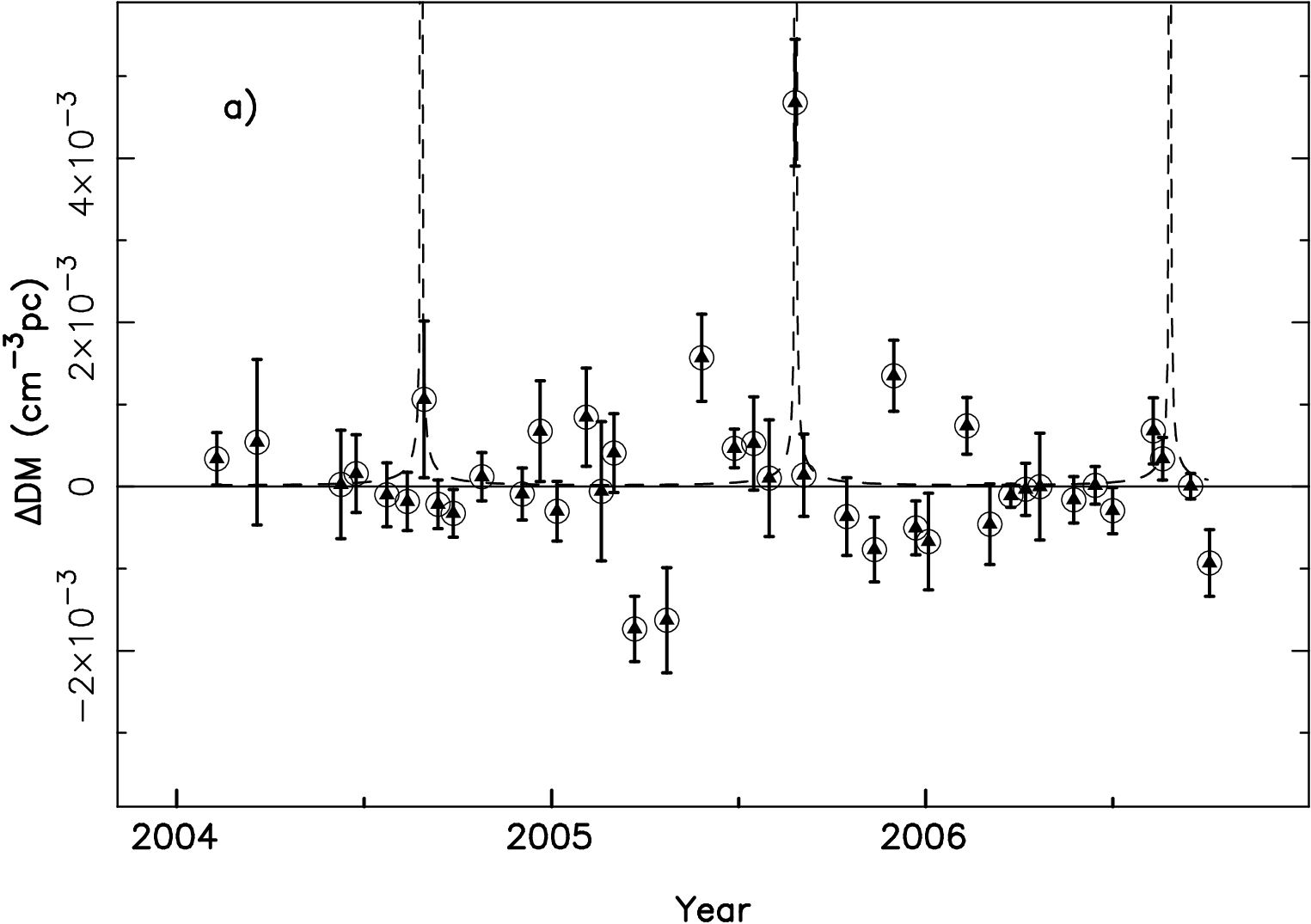}
    \includegraphics[angle=-90, width=82mm]{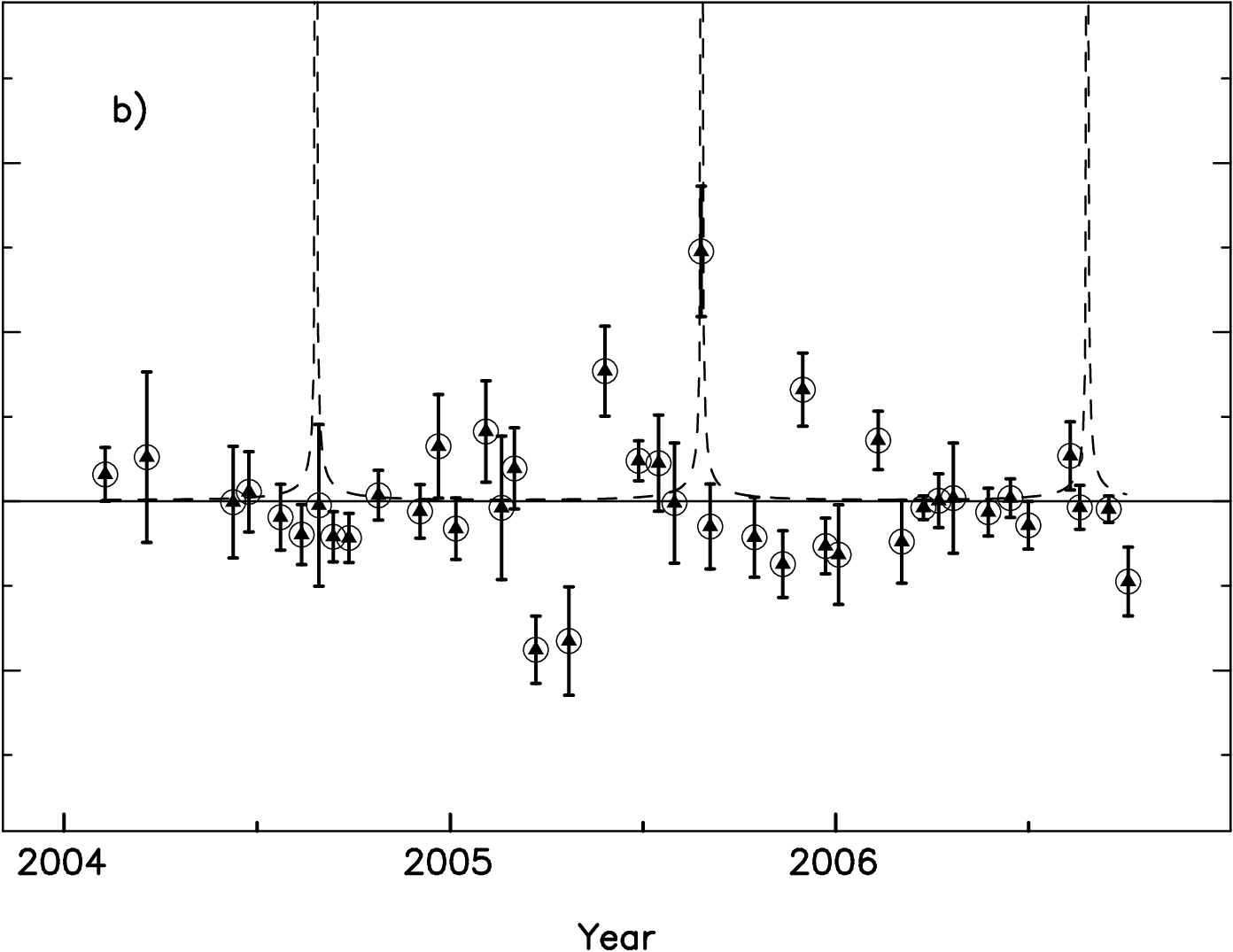}
  \end{tabular}
  \caption{DM variations of PSR~J1022$+$1001. The dashed curve is the
\textsc{tempo2} modelled $\rm DM_{\odot}$ values. Panel a shows the DM
variations with no correction for the solar wind. Panel b shows the DM
variations after correction by the model used in \textsc{tempo2}.}
  \label{fg:sw}
\end{figure*}

For PSRs~J1045$-$4509, J1824$-$2452 and J1909$-$3744 we have added an
indication of the inner time-scale (see \S5.3).  For PSRs~J1939$+$2134
and J1824$-$2452, we also overlay a dotted line which is the structure
function from earlier work (see \S5.3).

Our $\tau_d$ values for PSR~J0437$-$4715 can be compared with the
work of \citet{sgs06b} who scaled the observations of \citet{jnk98}
and \citet{gg00} by a large factor, assuming that the scintillation
index was much smaller than unity, although the scintillation index
was not reported by the original observers. \citet{sgs06b} deduced a
phase structure function which is two orders of magnitude lower than
ours in the vicinity of $\tau_d$. 

\subsection{Summary of results}

For all the pulsars we find that the measured structure functions lie
above the lower error bound of the Kolmogorov model.  Two,
PSRs~J0437$-$4715 and J1939$+$2134, fit the Kolmogorov model well. Two,
PSRs~J1045$-$4509 and J1909$-$3744, are clearly inconsistent with a
pure Kolmogorov power law, requiring a large inner scale. One,
PSR~J1824$-$2452, has few good $\tau_d$ measurements, but may well have
$\tau_i > \tau_d$. The remaining 15 pulsars are dominated by white
noise at small lags, and for five of these we cannot constrain the
slope of the underlying power-law spectrum. Two pulsars
(PSRs~J1744$-$1134, J1857$+$0943) could not be classified on the basis
of our measurements, but appear to be Kolmogorov on the basis of
previously published $\mbox {dDM}/ \mbox{d}t$ values. For five pulsars
(PSRs~J0613$-$0200, J1600$-$3053, J1643$-$1224, J1713$+$0747 and
J1732$-$5049) the structure functions fall below the quadratic model
at large time lags, strongly suggesting that the underlying spectrum
is Kolmogorov. The final three pulsars (PSRs~J1603$-$7202,
J1730$-$2304 and J2124$-$3358) appear to follow the quadratic model at
large lags. However it should be realised that the structure functions
at large lags are relatively poorly estimated and this separation of
the pulsars into different categories is not perfectly clear.

The results described above lead us to propose that the structure
functions for all our pulsars contain an ISM component that is either
a pure Kolmogorov power-law or a Kolmogorov power-law with a large
inner scale.

\section{Discussion of DM measurements}

\subsection{Comparison with earlier work}

 Much of the earlier work has concentrated on measuring (and
 modelling) $\mbox {dDM}/ \mbox{d}t $ values \citep[e.g.][]{bhvf93,hlk+04}.  
 For comparison with earlier work we have listed in
 Table~\ref{tb:DM} the slope of the best-fitting straight line across the
 entire data-set for each of our pulsars, $\mbox {dDM}/
 \mbox{d}t$. Our values generally do not agree with the previously
 published values. However, for our data-sets, a single $\mbox {dDM}/
 \mbox{d}t$ value models the observed $\Delta \mbox{DM}(t)$ values well only
 for a few pulsars (PSRs~J1045$-$4509, J1824$-$2452 and J1909$-$3744)
 and the $\mbox {dDM}/ \mbox{d}t$ values for other pulsars are
 misleading. For instance, for PSR~J0437$-$4715 our results indicate
 that the DM evolution for this pulsar can roughly be described using
 three $\mbox {dDM}/ \mbox{d}t$ values: prior to MJD~53400 $\mbox
 {dDM}/\mbox{d}t = (-2.98\pm0.07)\times10^{-4}\rm cm^{-3}pc\,yr^{-1}$,
 between MJD~53400 and 53700, $\mbox {dDM}/\mbox{d}t =
 (6.2\pm0.2)\times10^{-4}\rm cm^{-3}pc\,yr^{-1}$ and subsequently
 $\mbox {dDM}/\mbox{d}t = (3\pm1)\times10^{-5}\rm
 cm^{-3}pc\,yr^{-1}$. Clearly, the $\Delta \mbox{DM}(t)$ values are better
 described using the structure function.

\subsection{The solar wind}\label{sec:swind}

 The solar wind leads to a significant change in DM for pulsars close
 to the ecliptic plane during close approaches of the line of sight to
 the pulsar with the Sun. According to the \textsc{tempo2} model
 described in \S2, an unmodelled solar wind contributes
 $\sim$100\,ns at 1400 MHz for sources within $60^\circ$ of the Sun
 and $\sim$1\,$\mu$s within $7^\circ$.  It is therefore clear that
 corrections are necessary for 18 out of our 20 PPTA pulsars. The
 correction can potentially be made by modelling the solar wind or by
 directly measuring the DM to sufficient accuracy using multiple
 frequency observations.

 The solar wind varies with time and position. An overview of the
 relevant solar physics can be found in \citet{sch06}. Most of the
 variations in the solar wind are ascribed to a slowly changing
 spatial pattern that rotates with a 27-day period. In addition global
 transients, called coronal mass ejections (CME), occur every few
 days.  The chances of a given observation, which typically has a
 duration of 30-60\,min, encountering a CME are only a few percent, so
 these are not the most important effects. The quasi-static spatial
 pattern is roughly bimodal, the ``slow solar wind'' with high
 electron density is concentrated within about $\pm20^\circ$
 \citep{mbf+00} and the ``fast solar wind'' with lower electron
 densities at higher latitudes. The density difference is a factor of
 four at 1 AU and increases near the Sun. The correction necessary for
 a given observation can then vary by a factor of four depending on
 how much of the line of sight is in the slow versus the fast wind.

 Corrections for the solar wind have been attempted in both
 \textsc{tempo} and \textsc{tempo2}.  Both programs implement
 constant-density spherically symmetric models. \textsc{Tempo} uses a
 high density model where the electron density at 1\,AU, $n_e(\rm 1
 AU) = 10$\,cm$^{-3}$, whereas \textsc{tempo2} has a lower density
 model of $n_e(\rm 1 AU) = 4$\,cm$^{-3}$. \citet{sns+05} and
 \citet{lkn+06} used an identical spherically symmetric model, but
 fitted for the electron density.  They obtained $n_e(\rm 1 AU) =
 5\pm4$\,cm$^{-3}$ and $n_e(\rm 1 AU) = 6.9\pm2.1$\,cm$^{-3}$
 respectively.  However, it is not possible for a spherically
 symmetric model to correct the average timing residual due to the
 large difference in density between the fast and slow winds. This is
 clearly demonstrated by our PSR~J1022$+$1001 observations. The left
 panel in Figure~\ref{fg:sw} shows the DM variations of this pulsar
 without correcting for the solar wind and gives the correction from
 the \textsc{tempo2} model as a dashed line.  The right panel shows
 the DM variations after correction using the \textsc{tempo2} model. During
 the year 2004, the \textsc{tempo2} model did accurately correct the
 effect. The original \textsc{tempo} model which uses a larger
 electron density overcorrects these observations. The opposite occurs
 during 2005, when the \textsc{tempo2} model under-corrects the
 observations whereas the original \textsc{tempo} models the solar
 wind well.

 It is possible to use coronal measurements to improve our correction
 by estimating which parts of the line of sight are in the fast and
 which in the slow wind. This can be demonstrated with our
 PSR~J1022$+$1001 data, but it is not yet clear whether using an updated
 model to correct the observations improves on simply measuring the
 excess DM using multi-frequency observations. This work will be
 reported in a future publication.

\subsection{Spectrum of the ISM}

All but six pulsars are consistent with a Kolmogorov fluctuation
spectrum with an inner time-scale smaller than $\tau_d$. The clearest
examples are PSRs~J0437$-$4715 and J1939$+$2134.  The structure
function for PSR~J0437$-$4715 lies slightly above the upper bound of
the Kolmogorov model fit to the $\tau_d$ data.  However, if the
$\tau_d$ data were divided by a factor of 1.35 they would be
consistent. A line with this shift is shown in Figure~\ref{fg:sf1}a
through the $\tau_d$ data. Given the large scatter in $\tau_d$ we
consider that there is no evidence for an inner scale, nor do we see a
need to rescale $D_\phi(\tau_d)$ as did Smirnova et al. (2006). The
structure function at the largest time-scales is currently consistent
with a Kolmogorov process, but there is an indication that the
structure function may be flattening at these scales, as if an outer
scale around 60\,AU were present.  We do not expect such a small outer
scale, but it is not impossible if the turbulence has inhomogeneities
of this order. Such structures could be caused, for example, by shear
instabilities or due to a large-scale damping mechanism such as
ion-neutral damping.  The presence or absence of this flattening will
become clearer in a few years, as we accumulate more observations of
this pulsar.

 There have been several analyses of the structure function for
 PSR~J1939$+$2134 \citep{cwd+90,rdb+06}. The recent
 result of \cite{rdb+06} is overlaid on our structure function in
 Figure~\ref{fg:sf2}d. They fitted for the power-law exponent and
 obtained $\alpha = 1.66\pm0.04$. They also compare their $D_{\rm DM}$
 with a single $\tau_d$ measurement of 180\,s. This comparison
 suggests an inner scale of $1.3 \times 10^9$\,m. Our $D_{\rm DM}$
 observations are slightly above the Kolmogorov model fit to the
 $\tau_d$ observations.  However, as with PSR~J0437$-$4715 there is a
 large scatter in $\tau_d$. A shift of 1.43, shown as a solid line
 through the data in Figure~\ref{fg:sf2}d, would make the $D_{\rm DM}$
 consistent. Thus we believe that the case for an inner time-scale
 greater than $\tau_d$ is weak for this pulsar.

The structure functions for PSRs~J1045$-$4509 and J1909$-$3744 both
lie well above the upper bound of the Kolmogorov spectrum and require
an inner scale which is much larger than $V\tau_d$. We can identify a
break in the structure function for PSR~J1045$-$4509 which suggests an
inner time-scale of about 12\,d. The observations of PSR~J1909$-$3744
can set a lower bound on the inner time-scale of $500$\,d. To our
knowledge these are the first observations of such large inner
scales. In order to estimate the corresponding spatial scales we use
$V \approx 3.85\times 10^4(\nu_d D)^{1/2}/(\nu \tau_d)$ for a thin
screen \citep{grl94}. In both cases the inner scales of 0.7 and 20\,AU
are comparable to or larger than the refractive scales of 0.38 and
0.19\,AU respectively \footnote{The refractive time-scales are found
using $t_r \approx 2\nu/\nu_d \tau_d$ \citep{ric90}}.  These scales
are so large that they can only be compared with ion-neutral damping
scales. Ion-cyclotron damping at such scales would require absurdly
low densities (see Equation~\ref{eqn:li}).

Measurements of the structure function for PSR~J1824$-$2452 have been
published by \citet{cl97a} and shown to be consistent with a
Kolmogorov process. In Figure~\ref{fg:sf2}b we overlay the earlier
structure function (dotted line) with our results. The two analyses
are consistent.  Our $\tau_d$ estimates suggest that $\tau_i >
\tau_d$, but because there are few measurements of $\tau_d$ and the
inferred $\tau_i$ is not as a large as for PSRs~J1045$-$4509 and
J1909$-$3744, the evidence for a large inner scale is weak.

In analysing the structure functions we have assumed that the line-of-sight 
velocity is constant. However the true velocity is a vector sum,
weighted by the distance of the scattering plasma, of the pulsar
proper motion, the orbital velocity for binary pulsars, the velocity
of the plasma with respect to the local standard of rest and the
orbital velocity of the earth \citep{rcm00}. For many
of our pulsars these effects are important. Thirteen of our pulsars
are in binary systems, and in five of those the orbital velocity is
comparable with the proper motion. In all five of these plus another
four non-binary pulsars the Earth's orbital velocity is also
comparable. For these pulsars the magnitude and direction of the
velocity can change significantly, both on a time-scale of days, and
annually.

Diffractive observations are made on a time-scale which is short
compared to the orbital periods, so such observations are affected by
the instantaneous vector sum of velocities. DM variations are normally
averaged over times longer than the typical binary periods in our
sample, but shorter than a year. Thus these measurements are not
affected by the binary orbital velocity. On average the diffractive
observations see a higher velocity than the DM observations which will
lead to the temporal structure function being flatter than the spatial
structure function. If the turbulence in the interstellar plasma is
anisotropic, and there is increasing evidence that such anisotropy is
common, then the apparent diffractive time-scale will depend strongly
on the direction of the velocity.

The net effect on our estimation of the structure function of DM is
not large, because four of our five best constrained observations have
relatively small velocity modulation. However, observations of the
solitary pulsar PSR~J1939$+$2134 are strongly modulated by the Earth's
orbital velocity. In this case both diffractive and DM observations
see the same time-varying velocity so the slope of the structure
function is not altered, but both observations will be ``noisier''
than expected. In fact, the $D_{\rm DM}(\tau)$ for this pulsar is noisier
than expected, suggesting that we are seeing the effect of annual
variations in velocity.  This effect is largest in PSR~J2145$-$0750,
for which the proper motion and orbital velocity are both similar to
the Earth's orbital velocity. This pulsar shows a wider spread in
$\tau_d$ than most. The structure function for this pulsar is
white-noise dominated, making it difficult to estimate its slope of
the structure function. The expected velocity modulation makes it even
more difficult, so even though the structure function appears to be
Kolmogorov on the basis of a single $\mbox {dDM}/ \mbox{d}t$ value,
more observations are required to confirm this.

In PSRs~J1045$-$4509 and J1909$-$3744 we observe intensity variations
at the diffractive time-scale.  However if the inner scales were
greater than the refractive scales as they appear to be, then the
structure function is quadratic and no intensity scintillation should
be observed\footnote{A pure quadratic structure function corresponds
to a linear phase gradient, which simply shifts the apparent position
of the source and does not change the intensity.}\citep{wan80}.  Thus
there must be an underlying Kolmogorov process which is roughly equal
in amplitude to the steep-spectrum process at $\tau_d$. 

This situation has been proposed theoretically (Zweibel, private
communication 2006). It can arise when the primary energy input to the
turbulence is at scales larger than the ion-neutral damping scale.
Energy will cascade down to the ion-neutral scale where most of it
will be absorbed. However some energy may `tunnel' through the damping
region to support a second Kolmogorov cascade at a lower level (in the
vicinity of the ion-neutral collision frequency, plasma waves are
evanescent). 

In the case of PSR~J1045$-$4509 the energy difference is about a
factor of 30. For PSR~J1909$-$3744 we can only say that the energy
difference is at least a factor of 30. This is a very intriguing
possibility which requires both observational and theoretical
followup.  The existence of steep spectra might be confirmed by VLBI
observations which should show an rms position wander of
$\lambda/(2\pi V\tau_d)$ on a time-scale of the inner time-scale
(where $\lambda$ is the wavelength). The baselines needed for a 1\,rad
rms phase difference are $V\tau_d$, where $\tau_d$ scales as
$\nu^{1.2}$. This is about 8000\,km for PSR~J1045$-$4509 and 6000\,km
for PSR~J1824$-$2452 at 1400\,MHz. It would be much more difficult to
measure PSR~J1909$-$3744 as the baseline, even at 327\,MHz, would be
50000\,km and the time-scale for position wander would be greater than
500\,d.

\subsection{White noise}

Twelve of our pulsars show a well-defined flattening of the structure
functions at small lags which indicates the presence of a white-noise
process that is substantially greater than the measurement
error. Although this has not been discussed in the context of DM
measurements before, it is a well-known anomaly in TOA measurements.
Observers have often rescaled their measurement error estimates to
match this white noise, but it is not clear that the additional white
noise is due to measurement error. It could be due to a process
intrinsic to the pulsar, unexplained calibration issues, or to
diffractive TOA noise. The latter process will have the same time
scale as diffractive intensity scintillation and is highly correlated
with the intensity scintillation. Its rms is of the order of $\tau_0 =
1/(2\pi\nu_d)$.  Although this phenomena has not been well-studied, it
has been observed directly \citep{lrc98} and discussed theoretically
\citep{rnb86}.  We have compared the observed white noise rms with
$\tau_0$ for each pulsar and find that this mechanism will need to be
considered for four of the PPTA pulsars: PSRs~J1045$-$4509,
J1600$-$3053, J1643$-$1224 and J1939$+$2134.  Since this mechanism is
correlated with intensity it may be possible to use intensity
measurements to correct it. It is likely that most of the white noise
is related to system calibration errors as it is known, at least for
some pulsars, to depend on the observing frequency and backend
instruments.

\section{Correction of residuals for DM variations}\label{sec:ht}

During the design of the PPTA project it was realised that DM
variations would be an important source of timing noise and, unless
corrected, would obscure the signature of many interesting phenomena
such as gravitational waves. Initial expectations were that
observations using the dual-band receiver would be used to determine
$\Delta \mbox{DM}(t)$ which, in turn, would be used to correct the
20\,cm timing residuals. Of course, the measurements of DM include a
white-noise component discussed earlier and hence, the correction for
the ``red'' DM variations adds white noise. Smoothing the DM data
before making the correction will reduce the white noise more than the
DM noise. However, choosing the optimal smoothing is non-trivial.

The problem is that the timing model includes numerous terms such as
parallax, position, proper motion, period and period derivative that
absorb some of the residuals due to DM variations. Fitting the timing
model to the residuals can substantially reduce the effect of DM
variations, but causes significant errors in the fitted parameters. We
cannot use the post-fit rms timing residual as a goodness measure,
since it would not change at all after correction if the DM effect has
been totally absorbed in the fitted parameters. Accordingly we have
calculated the optimal smoothing factor using a simple analytical
model which requires equally spaced observations. We have also
simulated the observations at the actual sampling intervals with known
model parameters and determined the optimal smoothing for the
simulated data without having to fit a timing model. This work is
outlined in Appendix C.

\begin{figure}
  \centering
    \includegraphics[angle=-90, width=85mm]{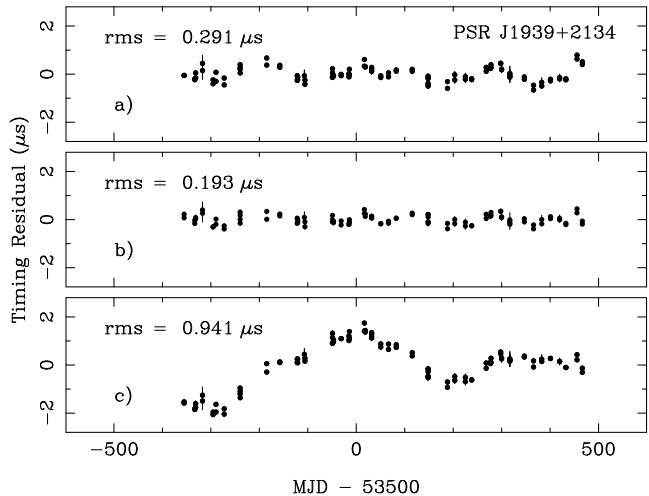}
  \caption{Timing residuals for PSR~J1939$+$2134. The upper panel (a)
  shows the timing residuals before DM correction, panel (b) contains
  the timing residuals after correcting for the DM variations using a
  71 day smoothing. Panel (c) shows the residuals obtained using the
  corrected parameters with the uncorrected TOAs.}
  \label{fg:tres}
\end{figure}

An example of the correction process is shown in Figure~\ref{fg:tres}. We have
shown the post-fit residuals for PSR~J1939$+$2134 before correction
(with an rms timing residual of $0.291$\,$\mu$s), and after correction
(rms of $0.193$\,$\mu$s). We can assume that the timing model after
correction is much more accurate than before correction. Therefore we
have plotted the residuals of the uncorrected data using the more
accurate corrected model resulting in an rms of $0.941$\,$\mu$s.  The
total proper motion, pulse-frequency, and frequency derivative were
changed during the correction process by 7$\sigma$, 28$\sigma$ and
28$\sigma$ respectively.

Figure~\ref{fg:tres} shows clearly the necessity of correction for DM
variations, and it also shows how fitting a timing model can
spuriously remove a ``red'' process. For instance, if the TOA
variations included the signature of a gravitational wave which
resembled the DM variations (both are expected to have a steep,
``red'' signature) then fitting the timing model would also have
removed most of the gravitational wave signature. Fortunately as the
duration of the timing data increases it becomes harder for the timing
model to emulate either the DM variations or the signature of
gravitational waves. This is because the terms related to the motion
of the Earth have annual or semiannual periods so they are not as
effective at removing longer period variations.

 The process described above has also been applied to five other
pulsars for which DM fluctuations are important. For each of these
pulsars, the theoretical smoothing time for uniformly sampled data and
the actual optimal smoothing time from the simulation are given in the
first two columns of Table~\ref{tb:ht}. The rms timing residuals
corresponding to the three panels of Figure~\ref{fg:tres},
``original'', ``corrected'', and ``true uncorrected'' (where the
corrected pulsar parameters are used to model the uncorrected TOAs),
are tabulated in the last three columns of the table. There is a
correlation between the improvement of the residuals and the slope of
$\Delta \mbox{DM}(t)$. The pulsars with the least slope,
PSRs~J0437$-$4715 and J1939$+$2134, showed the most improvement in rms
after correction for the DM variations.  This is because a linear
slope can be corrected exactly, but spuriously, in the original
fitting.  The $\Delta \mbox{DM}(t)$ values for all except the last
pulsar in the table are dominated by the plasma contribution. These
all show significantly higher ``true uncorrected'' residuals
demonstrating that the correction process was important even though it
may not have significantly lowered the rms timing residuals. The last
pulsar, PSR~J1643$-$1224, is dominated by white noise and does not
show much improvement in residual, nor is the true uncorrected
residual much larger.  As the PPTA continues to collect more data, the
DM corrections will become increasingly important and will have to be
applied to more of the observed pulsars.

\begin{table}
\caption{Improvement of timing residuals after correction for the DM variations.}
\label{tb:ht}
\begin{center}
\begin{tabular}{cccccc}\hline
PSR Name & $T^{\rm theory}_{sm}$ & $T^{simu.}_{sm}$ & $\sigma_{\rm orig}$ & $\sigma_{\rm cor.}$ & $\sigma_{\rm uncor.}$ \\ 
 & (d) & (d) & ($\mu$s) & ($\mu$s) & ($\mu$s) \\ \hline
J1939$+$2134 & 100 & 71  &  0.291 & 0.193 & 0.941 \\
J1824$-$2452 & 54  & 51  &  0.937 & 0.883 & 4.111 \\
J0437$-$4715 & 243 & 91  &  0.396 & 0.316 & 0.509 \\
J1909$-$3744 & 163 & 211 &  0.192 & 0.186 & 0.605 \\
J1045$-$4509 & 116 & 201 &  3.862 & 3.800 & 9.386 \\
J1643$-$1224 & 281 & 361 &  2.770 & 2.732 & 3.200 \\
\hline
\end{tabular}
\end{center}
\end{table}

\section{Conclusions}\label{sect:conclusion}

We have shown that correction for the plasma delay is essential
for the purposes of the PPTA project and have developed an optimal way
of applying this correction. 

We also show that the spherically symmetric solar wind models
included in the pulsar timing packages \textsc{tempo} and
\textsc{tempo2} are of marginal value.  More sophisticated models may
be useful, especially in situations where it is difficult or
impossible to measure the DM variations directly.

We have analysed the observed DM variations and found that most are consistent
with a simple Kolmogorov model of interstellar turbulence with
dissipation at a relatively small scale such as would be caused by ion
cyclotron damping. However at least two of the 15 pulsars for which we
can estimate the spectral exponent require a steeper spectrum and
suggest strongly that ion-neutral collisions are important in damping
the turbulence spectrum at AU scales.

\section*{Acknowledgements}
X.~P.~Y. is supported by the National Natural Science Foundation
(NNSF) of China (10521001).  The data presented in this paper were
obtained as part of the Parkes Pulsar Timing Array project which is a
collaboration between the ATNF, Swinburne University and The
University of Texas, Brownsville. The Parkes radio telescope is part
of the Australia Telescope which is funded by the Commonwealth of
Australia for operation as a National Facility managed by CSIRO.

\appendix
\section{Calculation of the structure function}

Let $D_{\rm est}(\tau)$ be the estimated structure function for the DM
variations at time lag $\tau$.  As described in \S2, we assume that
the errors on the $\Delta {\rm DM}(t)$ values are independent,
Gaussian and have known, but different, standard deviations.  The bias
and error that these errors contribute to $D_{\rm est}(\tau)$ are
obtained by expanding each $D_{\rm est}(\tau)$ as

\begin{eqnarray}
D_{\rm est}(\tau) &=& \frac{1}{N_p}\left \{ \sum_{ij}[\Delta \mbox{DM}(i)-
                      \Delta \mbox{DM}(j)]^2\right . \nonumber \\ 
                  &+& \sum_{ij}[e(i)^2 +e(j)^2]\nonumber\\ 
                  &+& 2\sum_{ij}[e(i)-e(j)][\Delta \mbox{DM}(i)-
                      \Delta \mbox{DM}(j)]\nonumber\\ 
                  &-&\left. 2\sum_{ij}e(i)e(j) \right \},
\end{eqnarray}
where $N_p$ is the number of pairs, $\Delta \mbox{DM}(i)$ is the
$\Delta \mbox{DM}$ within the time lag ``bin'' $\tau$ , and $e(i)$ is
the corresponding error. The first term in this expansion is the
desired estimator and the other terms are uncorrelated errors. The
second term is the only error term that does not have zero mean
and so contributes a bias which must be calculated and subtracted from
$D_{\rm DM}(\tau)$.  The variances of each term are easily calculated and
summed to give the total variance in $D_{\rm DM}(\tau)$. So finally, the
calculated $D_{\rm DM}(\tau)$ is
\begin{eqnarray}
D_{\rm DM}(\tau)&=& \frac{1}{N_p}\left \{ \sum_{ij}[\Delta \mbox{DM}(i)-
                      \Delta \mbox{DM}(j)]^2\right .\nonumber\\ 
            &-&\left . \sum_{i}N_i\epsilon(i)^2\right \}
                    \end{eqnarray}
where $\epsilon(i)$ is the rms of the $e(i)$, $N_i$ is
the number of times that $\Delta \mbox{DM}(i)$ is used to calculate
the $D_{\rm DM}(\tau)$. The variance of the $D_{\rm DM}(\tau)$ is
\begin{eqnarray} 
\sigma^2_{D_{\rm DM}}(\tau)&=&\frac{1}{N_p^2}\left \{2\sum_{i}N_i^2\epsilon(i)^4\right . 
                   \nonumber\\
                &+& 4\sum_{i}\sum_{j}\epsilon(i)^2
                  [\Delta \mbox{DM}(i)-\Delta \mbox{DM}(j)]^2\nonumber\\ 
                &+&\left. 4\sum_{ij}\epsilon(i)^2\epsilon(j)^2\right \}.
\end{eqnarray}
We have developed a \textsc{tempo2} plug-in that is publicly available
(download and usage instructions are given in Appendix D) to carry out
these calculations.

\section{Calculation of the diffractive time-scale and bandwidth}

The structure function of DM variations can be related to the diffractive
time ($\tau_d$) using Equation~\ref{eq:sts}.  Normally, the parameters of
diffractive interstellar scintillation (time-scale $\tau_d$ and
decorrelation frequency scale $\nu_d$) are obtained using a two
dimensional auto-correlation function (ACF) of the dynamic spectrum
$S(\nu,t)$ as
\begin{eqnarray}
&&C(\Delta \nu,\Delta t)=\nonumber\\
&&\frac{1}{N_{p}(\Delta \nu,\Delta t)}
\sum_{\nu}\sum_{t}\Delta S(\nu,t)\Delta S(\nu+\Delta \nu,t+\Delta t)
\end{eqnarray}
where $N_p$ is the number of pairs. $\Delta S(\nu,t)=S(\nu,t)-\bar{S}$
where $S(\nu,t)$ is the flux density and $\bar{S}$ is the mean flux
density for the whole observation.  The diffractive parameters are
defined by $C(0,\tau_d)=C(0,0)/e$ and $C(\nu_d,0)=C(0,0)/2$. The
parameters $\tau_d$ and $\nu_d$ are obtained by fitting a 2
dimensional Gaussian to $C(\Delta \nu,\Delta t)$.  However we often
have an observed dynamic spectrum which is not much longer than
$\tau_d$ and wider than $\nu_d$.  
 
We use a method based on the structure function instead of
auto-correlation function to estimate the diffractive time-scale. In
our data, the ACF is biased because we have few scintles in the
dynamic spectra.  For such cases the structure function, defined as
$D(\Delta t)=[S(t)-S(t+\Delta t)]^2/N_p$, is a better estimator
because it does not require estimation of $\bar{S}$. If $C(\Delta t)$
exists, then $D(\Delta t)=2[C(0)-C(\Delta t)]$. We estimate $D(\Delta
t)$ as

\begin{eqnarray}
\widetilde{D}(\Delta t) = \frac{1}{N_{p}(\Delta
t)}\sum_{\nu}\sum_{t}\left[\Delta S(\nu,t)-\Delta S(\nu,t+\Delta
t)\right]^2.
\end{eqnarray}

Because there is receiver noise which is white, the measured structure
function is
\begin{eqnarray}
  D_m(\Delta t)=D(\Delta t)+D_w = 2C(0)-2C(\Delta t)+D_w(\Delta t)
\end{eqnarray}
where $D_w$ is the structure function for the white noise.
\begin{eqnarray}
  D_w(\Delta t)&=&2\sigma^2  (\Delta t > 0) \nonumber \\
               &=&0 (\Delta t = 0 ) 
\end{eqnarray}
where $\sigma$ is the standard deviation of $S(\nu,t)$ measured over
the entire data-span.

If we normalise the flux density, then $\bar{S}=1$. For our
observations, the diffractive scintillation is strong and the
observing time is much less than the refractive time-scale, so
$C(0)=1$.  From the Kolmogorov spectrum, $C(\Delta t)=\exp(-(\Delta t/\tau_d)^{5/3})$.

So finally we can write the measured structure function as
\begin{eqnarray}
D_m(\Delta t)=2[1-\exp(-(\Delta t/\tau_d)^{5/3})]+D_w(\Delta t)
\label{eq:sf}
\end{eqnarray}

The uncertainty $\sigma_{D_m}(\Delta t)$ is estimated as
$\sigma_{D_m}(\Delta t) =D_m(\Delta t)\sqrt{\Delta t/T_{o}}$, where
$T_{o}$ is the observation time.  We choose the equal log time
interval points to fit because when $\Delta t$ is large, the points
are not independent.  Then we can fit the parameters $\tau_d$ and
$D_w$ in Equation~\ref{eq:sf} to obtain the diffractive time-scale
$\tau_d$.

A similar analysis can be used to obtain the diffractive bandwidth,
$\nu_d$.  However, in contrast to the determination of $\tau_d$, we
do not know the form of $C(\Delta \nu)$ and, hence, it is not possible to fit
for $\nu_d$ and $D_w(\Delta \nu)$.  However, $D_w(\Delta \nu) =
2\sigma^2 \approx D_m(\Delta \nu_{m})$, where $\Delta \nu_{m}$ is the
minimum frequency lag (in our data, it is typically 0.5\,MHz). 
$D_w(\Delta \nu)$ is the bias term which must be subtracted. This
leads to the structure function being,
\begin{eqnarray}
  D_s(\Delta \nu)= D_m(\Delta \nu) - D_m(\Delta \nu_m) = 2C(0)-2C(\Delta \nu).
\end{eqnarray}
After normalisation ($C(0)=1$) and according to the definition of
$\nu_d$ ($C(\nu_d)=C(0)/2=1/2$), we can obtain $\nu_d$ when
$D_s(\Delta \nu)=1$.

\section{Calculating the optimal smoothing time}

 Let $t_{g1}$ and $t_{g2}$ be idealised TOAs that are affected by
 neither noise nor DM variations. These TOAs correspond to frequencies
 $\nu_1$ and $\nu_2$ respectively where $\nu_1 > \nu_2$. Similarly,
 $t_{g1o}$ and $t_{g2o}$ are observed TOAs at these frequencies which
 have been affected by noise and DM variations.  So the observed TOAs
 are given by

 \begin{eqnarray}
  t_{gio}=t_{gi}+n_i(t)+\frac{\Delta \mbox{DM}(t)}{K}\nu_i^{-2}
 \label{eq:to}
 \end{eqnarray}
 where $n_i(t)$ is the noise at $\nu_i$ which is assumed to be white
 ($i=1,2$ for the two observations respectively). The measured
 estimate of $\mbox{DM}(t)$ is therefore given by
 \begin{eqnarray}
 \widetilde{\mbox{DM}}(t) &=&
 [t_{g2o}(t)-t_{g1o}(t)]\frac{K}{\nu_2^{-2}-\nu_1^{-2}}\nonumber\\ &=&
 (t_{g2}-t_{g1})\frac{K}{\nu_2^{-2}-\nu_1^{-2}}+\Delta \mbox{DM}(t)\nonumber
 \\ & &+[n_2(t)-n_1(t)]\frac{K}{\nu_2^{-2}-\nu_1^{-2}}.
 \label{eq:bcr}
 \end{eqnarray}
 After correcting for the DM variations, by subtracting the
 corresponding time offsets from $t_{g1o}$, we obtain a set of
 corrected TOAs, $t_{g1c}(t)$, which are given by,
 \begin{eqnarray}
  t_{g1c}(t)&=&t_{g1o}(t)-\frac{\widetilde{\mbox{DM}}(t)}{K}\nu_1^{-2}\nonumber\\
          &=&t_{g1} a_1-t_{g2} a_2+n_1(t)a_1-n_2(t)a_2
\label{eq:cr}
\end{eqnarray}
 where $a_1 = {\nu_2^{-2}}/({\nu_2^{-2}-\nu_1^{-2}})$ and $a_2 =
 {\nu_1^{-2}}/({\nu_2^{-2}-\nu_1^{-2}})$. Note that, since $a_1 - a_2 =
 1$, contributions to the timing residuals which are frequency
 independent and thus the same in $t_{g1}$ and $t_{g2}$, appear
 unchanged in $t_{g1c}$. This is an important property of the
 correction algorithm because interesting contributions such as the
 signature of gravitational waves, planets orbiting the pulsar, or
 ephemeris errors are unaltered by the correction.

 Comparing Equations~\ref{eq:to} and \ref{eq:cr} we see that the
 variance of the white noise has increased from $\sigma_{N1}^2$ to
 $\sigma_{N1}^2a_1^2+\sigma_{N2}^2a_2^2$ although the DM noise has
 been eliminated. We can improve the variance in $t_{g1c}$ by
 smoothing $\widetilde{\mbox{DM}}(t)$ before subtracting it from
 $t_{g1o}$, because smoothing reduces the white noise more than the
 $\mbox{DM}(t)$ variations.  The white variance in
 $\widetilde{\mbox{DM}}(t)$ is reduced by the smoothing number $N_s$.
 It is hard to calculate the effect of smoothing $\mbox{DM}(t)$
 analytically, but we found numerically that the variance of equally
 sampled $[\mbox{DM}(t) - {\rm{smoothed}}(\mbox{DM}(t))] \approx 0.5
 D_{\rm DM}(T_{sm}/2\pi)$ where $T_{sm} = (N_s - 1)\tau_0$ and $\tau_0$ is
 the sampling time (typically 15\,d).  However, the white noise in the
 smoothed $\widetilde{\mbox{DM}}(t)$ is partially correlated with that
 in $t_{g1o}$, so we have finally the variance in $t_{g1c}$
\begin{eqnarray}
 \sigma_{1csm}^2&=&\sigma_{N1}^2+\frac{(\sigma_{N1}^2+\sigma_{N2}^2)a_2^2}{N_s}
+\frac{2a_2\sigma_{N1}^2}{N_s}\nonumber \\
&&+\frac{1}{2}\left(\frac{1}{k\nu_1^2}\right)^2\left(\frac{N_s-1}{2\pi}\right)^{\alpha}D_{\rm DM}(\tau_0).
\end{eqnarray}
We minimise this numerically.  In fact, the data are not equally
sampled so we first interpolate the raw data on to an equally spaced
array before smoothing.  To check the effect of re-sampling we
compared the theory above with a simulation.  We realised 50 samples
of $\mbox{DM}(t)$ from a population matching $\mbox{DM}(\tau)$ with
the actual data sampling.  Then we interpolated the simulated data
onto an equally spaced array and found the $N_s$ which best corrected
for $\mbox{DM}(t)$.  In all cases the minimum is very broad so it
makes little difference whether one uses the theoretical or simulated
value of $N_s$.  However, the simulation is easy to implement and will
be correct even in the case of an unusual distribution of samples.

\section{Available software}

The \textsc{tempo2} software was designed to allow for easy addition
of new features and functionality in the form of plug-ins to the main
package.  During this work we have produced the following new plug-in
packages which are now available as part of the \textsc{tempo2}
distribution (full details are available from our web-site,
\url{http://www.atnf.csiro.au/research/pulsar/tempo2}):

\begin{itemize}
\item{\textsc{calcDM}: this plug-in contains the algorithms described
  in this paper. This plug-in allows the user to calculate and plot
  $\Delta \mbox{DM}(t)$ and obtain the corresponding structure function.}
\item{\textsc{sf}: calculates and plots the structure function of the
  timing residuals.}
\item{\textsc{simISM}: allows data-sets to be simulated in order to
study the effect of a Kolmogorov process on pulsar timing residuals.}
\end{itemize}

The program \textsc{diffTime} can be used to calculate $\tau_d$ and
$\nu_d$ for most pulsar observations.  This software is available as
part of the \textsc{PSRchive} software distribution
(\url{http://psrchive.sourceforge.net}).

Software to simulate the effect of refraction on $\tau_d$ estimates is
in the \textsc{sim 2.0} distribution from UCSD:
\url{http://typhoon.ucsd.edu/~coles/sim2.0/sim2.0.html}.

\end{document}